\documentclass[letterpaper,english,prd,preprint,nofootinbib,showpacs,preprintnumbers,fleqn,floatfix]{revtex4}
\usepackage[T1]{fontenc}
\usepackage[latin1]{inputenc}
\usepackage{array}
\usepackage{graphicx}
\usepackage{amssymb}

\makeatletter

\providecommand{\tabularnewline}{\\}

\usepackage{babel}
\makeatother
\begin{document}

\newcommand{\prescr}[1]{{}^{#1}}

\newcommand{\ra}{\rightarrow}

\newcommand{\alphas}{\alpha_{s}}

\newcommand\gsim{\mathrel{\rlap{\raise.4ex\hbox{$>$}} {\lower.6ex\hbox{$\sim$}}}}  \newcommand\lsim{\mathrel{\rlap{\raise.4ex\hbox{$<$}} {\lower.6ex\hbox{$\sim$}}}}  

\newcommand\etmiss{E_{T}\hspace{-13pt}/\hspace{8pt}}

\newcommand{\mHQ}{m_{\mathcal Q}}

\newcommand{\qHQ}{{\mathcal Q}}

\pacs{12.15.Ji, 12.38 Cy, 13.85.Qk}

\preprint{hep-ph/0509023}

\preprint{ANL-HEP-PR-05-73}

\title{Heavy-flavor effects in soft gluon resummation for electroweak boson
production at hadron colliders }

\author{Stefan Berge,$^{1}$%
\footnote{E-mail: berge@mail.physics.smu.edu%
} Pavel M. Nadolsky,$^{2}$%
\footnote{E-mail: nadolsky@hep.anl.gov%
} and Fredrick I. Olness,$^{1}$%
\footnote{E-mail: olness@smu.edu%
}}

\affiliation{$^{1}$Department of Physics, Southern Methodist University, Dallas,
Texas 75275-0175, U.S.A.\\
$^{2}$High Energy Physics Division, Argonne National Laboratory,\\
Argonne, IL 60439-4815, U.S.A. }

\begin{abstract}
We evaluate the impact of heavy-quark masses on transverse momentum
($q_{T}$) distributions of $W$, $Z$, and supersymmetric neutral
Higgs bosons at the Tevatron and LHC. The masses of charm and bottom
quarks act as non-negligible momentum scales at small $q_{T}$ and
affect resummation of soft and collinear radiation. We point out inconsistencies
in the treatment of heavy-flavor channels at small $q_{T}$ in massless
and fixed-flavor number factorization schemes, and formulate small-$q_{T}$
resummation in a general-mass variable flavor number factorization
scheme. The improved treatment of the quark mass dependence leads
to non-negligible effects in precision measurements of the $W$ boson
mass at the LHC and may cause observable modifications in production
of Higgs bosons and other particles in heavy-quark scattering.

\end{abstract}

\date{\today{}}

\maketitle

\section{Introduction\label{sec:Introduction}}

Key electroweak observables of the standard model (SM), such as the
mass and width of $W$ bosons, will be measured with high precision
in production of $W^{\pm}$ and $Z^{0}$ bosons in hadron collisions
at the Fermilab (Tevatron) and Large Hadron Collider at CERN (LHC).
The uncertainty in the $W$ boson mass $M_{W}$ will be reduced to
30-40 MeV per experiment in the Tevatron Run-2 and to about 15 MeV
at the LHC~\cite{Brock:1999ep,Haywood:1999qg}{\large .} At this
level of accuracy, the theory framework must incorporate effects were
dismissed in less precise analyses. 

One such feature is the dependence on masses of quarks, which is frequently
neglected in hard-scattering reactions. While the massless approximation
is clearly appropriate in channels involving only light quarks ($u$,
$d$, and perhaps $s$), it may be less adequate when heavy quarks
($c$ and $b$) are involved, particularly when the cross section
depends on a small momentum scale close to the mass $\mHQ$ of the
heavy quark. The distributions of heavy electroweak bosons over their
transverse momenta $q_{T}$ may be sensitive to the masses of $c$
and $b$ quarks, given that the transverse momenta of most bosons
(of order, or less than a few GeV) are comparable to $\mHQ$. The
impact of the quark masses on the $q_{T}$ distributions of bosons
with invariant masses $Q\geq M_{W}$ is suppressed by properties of
soft parton radiation, as demonstrated below. Nonetheless, the quark
mass effects are relevant in high-precision studies, such as the measurement
of $M_{W}$, or in reactions dominated by scattering of heavy quarks
(particularly bottom quarks), such as Higgs boson production in $b\bar{b}$
annihilation. 

In the present paper, we quantify the effects of the heavy-quark masses
on $q_{T}$ distributions in Drell-Yan production of $W^{\pm}$,~$Z^{0}$,
and Higgs bosons. We evaluate the associated impact on the $W$~boson
mass measurement and demonstrate the importance of the quark-mass
corrections in the processes initiated by bottom quarks, like $b\bar{b}\rightarrow\mbox{Higgs}$.
This latter channel can have a large cross section in supersymmetric
extensions of the standard model~\cite{Carena:2002es,Spira:1997dg,Assamagan:2004mu}.

The description of the heavy-quark scattering is complex because of
the presence of the heavy-quark mass scale $\mHQ$, in addition to
the boson's transverse momentum~$q_{T}$ and its virtuality~$Q$.
Two popular factorization schemes (fixed-flavor number (FFN) scheme~\cite{Gluck:1982cp,Gluck:1988uk,Nason:1989zy,Laenen:1992cc,Laenen:1993zk,Laenen:1993xs}
and zero-mass variable flavor number (ZM-VFN) scheme) were applied
recently to calculate various observables in the $b$-quark channels~\cite{Harlander:2003ai,Campbell:2002zm,Maltoni:2003pn,Dawson:2003kb,Dawson:2004sh,Dawson:2005vi,Dittmaier:2003ej,Campbell:2003dd,Maltoni:2005wd}.
However, neither of the two schemes is entirely consistent when describing
logarithmic corrections in processes like $b\bar{b}\rightarrow H$
in the small-$q_{T}$ region. When $q_{T}$ is much smaller than $Q$,
the calculation of the $q_{T}$~distribution in perturbative quantum
chromodynamics (PQCD) must evaluate an all-order sum of large logarithms
$\ln(q_{T}^{2}/Q^{2})$, arising as a consequence of the recoil of
the electroweak bosons against unobserved hadrons of relatively low
energy or transverse momentum (soft and collinear hadrons). The resummation
of Drell-Yan $q_{T}$ distributions for massless initial-state quarks
has been developed in a variety of forms~\cite{Dokshitzer:1978dr,Parisi:1979se,Altarelli:1984pt,Collins:1985kg,Ellis:1998ii,Kulesza:2002rh,Ji:2004xq}.
In the present paper we use the Collins-Soper-Sterman formalism (CSS)~\cite{Collins:1985kg},
which excellently describes the available $q_{T}$ data from fixed-target
Drell-Yan experiments and $Z^{0}$ boson production at the Tevatron.
When applied to heavy-quark scattering, small-$q_{T}$ resummation
indispensably involves logarithms depending on both the transverse
momentum, $\ln(q_{T}^{2}/Q^{2})$, and heavy-quark mass, $\ln(\mu_{F}^{2}/\mHQ^{2})$.
To correctly treat both types of large contributions, Ref.~\cite{Nadolsky:2002jr}
extended the CSS formalism to a general-mass variable flavor number
(GM-VFN) scheme~\cite{Collins:1998rz}. The extended resummation
formalism was then applied to describe heavy-flavor production in
deep inelastic scattering (DIS). We review the $q_{T}$ resummation
formalism for heavy quarks in Section~\ref{sec:Formalism}. 

The present study was motivated by the results of Ref.~\cite{Nadolsky:2002jr},
where the correct treatment of the heavy-quark masses was found to
substantially change the differential cross sections for heavy quark
production in DIS at small $q_{T}\approx\mHQ$. In Section~\ref{sec:Numerical-results},
we extend that study to examine the heavy-flavor effects on $q_{T}$
distributions of $W^{\pm}$, $Z^{0}$, and Higgs bosons produced at
the Tevatron and LHC. To get a first idea about the magnitude of the
mass effects, we consider fractional contributions of reaction channels
with initial-state $c$ and $b$ quarks to the entire production rate
(see Table~\ref{tab:WZfraction}). At the Tevatron, the heavy-quark
channels contribute only $8\%$ ($3\%$) of the inclusive cross section
in $W^{\pm}$ ($Z^{0}$) boson production, and consequently the quark
masses can be usually neglected. At the LHC, the heavy-quark channels
add up to $22\%$ in $W^{+}$ and $31\%$ in $W^{-}$ boson production.
Modifications caused by the heavy-quark masses at the LHC are comparable
to the other uncertainties and must be considered in precision measurements
of $M_{W}$. We then turn to Higgs boson production in the $b\bar{b}\rightarrow H$
channel. We show that the nonzero mass of the bottom quark substantially
modifies the small-$q_{T}$ Higgs distribution. Furthermore, we note
that other processes with initial-state heavy quarks, like $bg\rightarrow Z^{0}b,\, H^{-}t,\, H^{+}\bar{t},\, H^{0}b$,
etc. will be affected by the heavy-quark masses in a similar way.
We summarize the results of this study in Section~\ref{sec:Conclusion}.

\section{\label{sec:Formalism}Transverse momentum resummation for massive
quarks}

\subsection{General form of the resummed form factor\label{sub:GeneralFormW}}

In this section, we define the essential elements for the transverse
momentum resummation in the presence of massive quarks. The resummed
differential cross section for the inclusive production of heavy electroweak
bosons in scattering of initial-state hadrons $A$ and $B$ takes
the form~\cite{Collins:1985kg}\begin{equation}
\frac{d\sigma}{dQ^{2}dydq_{T}^{2}}=\int_{0}^{\infty}\frac{bdb}{2\pi}\, J_{0}(q_{T}b)\,\widetilde{W}(b,Q,x_{A},x_{B},\{\mHQ\})\,\,+\,\, Y(q_{T},Q,y,\{\mHQ\}),\label{WYDY}\end{equation}
 where $y=\left(1/2\right)\ln\left[(E+p_{z})/(E-p_{z})\right]$ is
the rapidity of the vector boson, $x_{{A,B}}\equiv Qe^{\pm y}/\sqrt{S}$
are the Born-level partonic momentum fractions, $S$ is the square
of the center-of-mass energy of the collider, and $J_{0}(q_{T}b)$
is the Bessel function. The integral is the Fourier-Bessel transform
of a form factor $\widetilde{W}(b,Q,x_{{A}},x_{{B}},\{ m_{q}\}$)
in impact parameter ($b$) space, which contains the all-order sum
of the logarithms $\alpha_{s}^{n}\,\ln^{m}(q_{T}^{2}/Q^{2})$. The
$b$-space form factor is given by\begin{equation}
\widetilde{W}(b,Q,x_{{A}},x_{{B}},\{\mHQ\})=\frac{\pi}{S}\,\sum_{j,k}\sigma_{jk}^{(0)}\, e^{-\mathcal{S}(b,Q,\{\mHQ\})}\,\,{\mathcal{\overline{P}}}_{j/A}(x_{A},b,\{\mHQ\})\,\,{\mathcal{\overline{P}}}_{k/B}(x_{{B}},b,\{\mHQ\}),\label{WCSS}\end{equation}
 where the summation is performed over the relevant parton flavors
$j$ and $k$. Here, $\sigma_{jk}^{(0)}$ is the product of the Born-level
prefactors, $e^{-{\mathcal{S}}(b,Q,\{\mHQ\})}$ is an exponential
of the Sudakov integral \begin{eqnarray}
{\mathcal{S}}(b,Q,\{\mHQ\})\equiv\int_{b_{0}^{2}/b^{2}}^{Q^{2}}\frac{d\bar{\mu}^{2}}{\bar{\mu}^{2}}\biggl[{\mathcal{A}}(\alpha_{s}(\bar{\mu}),\{\mHQ\})\,\mathrm{ln}\biggl(\frac{Q^{2}}{\bar{\mu}^{2}}\biggr)+{\mathcal{B}}(\alpha_{s}(\bar{\mu}),\{\mHQ\})\biggr],\label{Sudakov}\end{eqnarray}
 and ${\mathcal{\overline{P}}}_{j/A}(x,b,\{\mHQ\})$ are the $b$-dependent
parton distributions. The constant factor $b_{0}\equiv2e^{-\gamma_{E}}\approx1.123$
appears in several places when a momentum scale is constructed from
the impact parameter, as in the lower limit $b_{0}^{2}/b^{2}$ of
integration in Eq.~(\ref{Sudakov}). The Sudakov exponential and
$b$-dependent parton densities resum contributions from soft and
collinear multi-parton radiation, respectively. In the perturbative
region ($b^{2}\ll\nolinebreak\Lambda_{QCD}^{-2}$), the distributions
${\mathcal{\overline{P}}}_{j/A}(x,b,\{\mHQ\})$ factorize as\begin{eqnarray}
\left.\overline{{\mathcal{P}}}_{j/A}(x,b,\{\mHQ\})\right|_{b^{2}\ll\Lambda_{QCD}^{-2}} & = & \sum_{a=g,u,d,...}\,\int_{x}^{1}\,\frac{d\xi}{\xi}\,{\mathcal{C}}_{j/a}(x/\xi,b,\{\mHQ\},\mu_{F})\, f_{a/A}(\xi,\mu_{F})\nonumber \\
 & \equiv & \left({\mathcal{C}}_{j/a}\otimes f_{a/A}\right)(x,b,\{\mHQ\},\mu_{F})\label{Pja}\end{eqnarray}
 into a sum of convolutions of the Wilson coefficient functions ${\mathcal{C}}_{j/a}(x,b,\{\mHQ\},\mu_{F})$
and $k_{T}$-integrated parton distributions $f_{a/A}(\xi,\mu_{F})$.
Note that the initial state $A$ can be a hadron as well as a parton,
in which case we denote the partonic initial state as $A^{\prime}$.

If quark masses $\{\mHQ\}$ are neglected in the coefficient function
${\mathcal{C}}_{j/a}(x,b,\{\mHQ\},\mu_{F})$, it depends on $b$ and
$\mu_{F}$ only through the logarithm $\ln(\mu_{F}b/b_{0})$:\begin{equation}
\lim_{\{ b\mHQ\}\rightarrow\{0\}}{\mathcal{C}}_{j/a}(x,b,\{\mHQ\},\mu_{F})={\mathcal{C}}_{j/a}(x,\ln\frac{b\mu_{F}}{b_{0}}).\label{smallbmHQ}\end{equation}
 For this reason, we set the factorization scale $\mu_{F}$ equal
to $b_{0}/b$ to prevent the large logarithms $\ln(\mu_{F}b/b_{0})$
from appearing in ${\mathcal{C}}_{j/a}$ in the limit $b\rightarrow0$. 

$Y(q_{T},Q,y,\{\mHQ\})$ in Eq.~(\ref{WYDY}) is the regular part,
defined as the difference of the fixed-order cross section and the
expansion of the Fourier-Bessel integral to the same order of $\alpha_{s}$;
it dominates at $q_{T}\sim Q$ and is small at $q_{T}\rightarrow0$.
The regular piece $Y$ and the dominant contributions in $\widetilde{W}$
can be calculated in PQCD, if $Q$ is sufficiently large. A small
nonperturbative component in $\widetilde{W}$ contributing at $q_{T}$
less than a few GeV can be approximated within one of the available
models. Our specific choice of the nonperturbative model is described
in Section~\ref{sub:Non-perturbative-contributions}.

\subsection{Extension to the case of massive quarks \label{sub:ExtensionToTheMassiveCase}}

Eqs.~(\ref{WYDY}) and~(\ref{WCSS}) indicate that $\widetilde{W}$
and $Y$ may explicitly depend on the masses of heavy quarks $\{\mHQ\},$
with $\qHQ=c,$ $b,$ and $t$. Masses of the light $u,$ $d,$ and
$s$ quarks ($m_{u,d,s}\lsim\Lambda_{QCD}$) are neglected by definition
in hard-scattering contributions and implicitly retained in the nonperturbative
functions, \emph{i.e.}, parton densities $f_{a/A}(x,\mu_{F})$ and
power corrections to $\widetilde{W}$ at large impact parameters $b>1$~$\mbox{GeV}^{-1}$.
Indeed, the nonperturbative dynamics described by $f_{a/A}(x,\mu_{F})$
and power corrections does depend on the light-quark masses, but this
dependence is not separated explicitly from the other nonperturbative
effects.

In contrast to the masses of $u,$ $d,$ and $s$ quarks, masses of
the heavy quarks must be retained in the hard-scattering terms in
some cases. Several viable options exist for the treatment of $\{\mHQ\}$,
depending on the energy and flavor composition of the scattering reaction.
In two common approaches, the heavy quark contributions to the PDF's
$f_{a/A}(x,\mu_{F})$ and other resummed functions are either neglected
or, alternatively, treated on the same footing as contributions from
the light quarks ($u,d,$ and $s$) above the respective heavy-quark
mass thresholds, often placed, for convenience, at $\mu_{F}=\mHQ$.
These choices correspond to the FFN and ZM-VFN factorization schemes,
respectively. The FFN scheme optimally organizes the perturbative
QCD series at energies of order of the heavy-quark mass $\mHQ$. The
ZM-VFN scheme is the best choice for inclusive (depending on one momentum
scale) distributions at energies much larger than $\mHQ$. It commonly
utilizes dimensional regularization to expose collinear singularities
of massless partonic matrix elements as $1/\epsilon^{p}$ poles in
$n=4-2\epsilon$ dimensions. 

Neither of the two schemes is entirely satisfactory when applied to
differential distributions depending on two or more momentum scales
of distinct magnitudes, as is the case of small-$q_{T}$ resummation
at large $Q$. At sufficiently large $\sqrt{S}$, the heavy-flavor
quarks are copiously produced by quasi-collinear splittings of gluons
along the directions of the initial-state hadrons. Such collinear
contributions must be resummed at $Q\gg\mHQ$ in the parton density
$f_{\qHQ/A}(x,\mu_{F})$ for the heavy quarks $\qHQ$, so that a VFN
factorization scheme is needed. If all scales (including $q_{T}$)
are of order $Q$, the heavy-quark mass $\mHQ\ll q_{T}\sim Q$ can
be neglected in the hard-scattering matrix elements, reducing the
result to the ZM-VFN scheme. But $\mHQ$ cannot be omitted at small
$q_{T}$ ($\mHQ\sim q_{T}\ll Q$), where it is not small compared
to the momentum of the soft and collinear radiation. 

In $b$-space, the form factor $\widetilde{W}(b,Q,x_{A},x_{B},\{\mHQ\})$
is not well-defined in the ZM-VFN scheme in the heavy-quark channels
at impact parameters $b>b_{0}/\mHQ$, corresponding to factorization
scales $\mu_{F}$ below the heavy-quark mass threshold, $\mu_{F}=b_{0}/b<\mHQ$.
For charm quarks with mass $m_{c}=1.3\mbox{\, GeV}$ and bottom quarks
with mass $m_{b}=4.5$~GeV, the ZM-VFN form factor $\widetilde{W}$
is not well-defined at $b>\nolinebreak0.86$ and $0.25\mbox{\, GeV}^{-1}$,
respectively. The problem lies with the heavy-quark Wilson coefficient
function ${\mathcal{C}}_{\qHQ/a}(x,b,\mHQ,\mu_{F})$, which is derived
by using the factorization relation of Eq.~(\ref{Pja}) with the
densities ${\mathcal{\overline{P}}}_{\qHQ/A^{\prime}}(x,b,\mHQ)$
and $f_{a/A^{\prime}}(x,\mu_{F})$ of heavy quarks in partonic initial
states $A^{\prime}=q,\qHQ,g$. At $\mu_{F}$ below the threshold,
the $k_{T}$-integrated heavy-quark density $f_{\qHQ/A^{\prime}}(x,\mu_{F})$
is set identically equal to zero, in accordance with the definition
of $f_{\qHQ/A^{\prime}}(x,\mu_{F})$ in the ZM-VFN scheme. Consequently
the collinear poles $1/\epsilon$ arising in the calculation of the
$b$-dependent density ${\mathcal{\overline{P}}}_{\qHQ/A^{\prime}}(x,b)$
for massless splittings $\qHQ\leftarrow A^{\prime}$ in $n\neq4$
dimensions cannot be canceled by $f_{a/A^{\prime}}(x,\mu_{F})$ at
such $\mu_{F}$. Eq\@.~(\ref{Pja}) then implies that the Wilson
coefficient function ${\mathcal{C}}_{\qHQ/a}(x,b,\mHQ,\mu_{F})$ may
be also infinite (contain the $1/\epsilon$ poles) below the heavy-quark
threshold. The infinity arises because the ZM-VFN scheme incorrectly
neglects the heavy-quark mass $\mHQ$ at energies $\mu_{F}=b_{0}/b$
of order or less than $\mHQ$. The solution to this problem is to
retain the dependence on $\mHQ$ at $b\gtrsim b_{0}/\mHQ,$ which
can be realized by formulating the CSS resummation in the GM-VFN factorization
scheme. 

The GM-VFN scheme \cite{Collins:1998rz} consistently implements the
heavy-quark masses at all energy scales $\mu_{F}$, and it reproduces
the FFN and ZM-VFN schemes in the limits $\mu_{F}\lesssim\mHQ$ and
$\mu_{F}\gg\mHQ$, respectively. Several versions of the GM-VFN scheme
have been developed in the past years~\cite{Aivazis:1994pi,Kniehl:1995em,Buza:1998wv,Thorne:1998ga,Thorne:1998uu,Cacciari:1998it,Chuvakin:1999nx,Kramer:2000hn,Tung:2001mv}.
A general procedure for the implementation of the GM-VFN scheme in
$q_{T}$ resummation was outlined in Ref.~\cite{Nadolsky:2002jr},
where an application of the CSS resummation in the simplified Aivazis-Collins-Olness-Tung
(S-ACOT) factorization scheme~\cite{Collins:1998rz,Kramer:2000hn}
was presented for production of heavy quarks at HERA.

The S-ACOT scheme is a variant of the GM-VFN scheme, which simplifies
computations by neglecting the heavy-quark masses in hard subgraphs
with incoming heavy quarks (flavor-excitation graphs), such as $\qHQ+g\rightarrow\qHQ+g$.
The heavy-quark masses are retained in the hard subgraphs with explicit
production of heavy quarks (flavor-creation graphs), such as $g+g\rightarrow\qHQ+\bar{\qHQ}$.
As demonstrated in Ref.~\cite{Nadolsky:2002jr}, application of the
S-ACOT rules to the CSS resummed cross section efficiently retains
the dependence on the heavy-quark masses where it is important, and
drops it where it is not essential.

In production of heavy gauge bosons ($Q\gg\mHQ$), the S-ACOT rules
allow us to drop the $\mHQ$ dependence at $q_{T}\gg\mHQ$ and keep
the essential $\mHQ$ dependence at $q_{T}\lesssim\mHQ$. We assume
that the heavy quarks are pairwise produced in perturbative splittings
of gluons. We neglect possible, but yet experimentally unconfirmed,
nonperturbative {}``intrinsic'' heavy-quark contributions to the
proton wavefunction~\cite{Brodsky:1980pb}. We neglect the mass dependence
entirely in the $Y$-term, as it is non-negligible only at $q_{T}\gg\mHQ$.
In the $\widetilde{W}$-term, we drop the $\mHQ$ dependence in all
flavor-excitation hard subgraphs and keep it in all flavor-creation
hard subgraphs. By this rule, $\mHQ$ is dropped in the perturbative
Sudakov form factor ${\mathcal{S}}$ and the coefficient functions
${\mathcal{C}}_{a/\qHQ}$ with the incoming heavy quarks, both of
which are described by the flavor-excitation Feynman graphs. We evaluate
${\mathcal{S}}$ up to ${\mathcal{O}}(\alpha_{s}^{2}/\pi^{2})$ and
${\mathcal{C}}_{a/\qHQ}$ up to ${\mathcal{O}}(\alpha_{s}/\pi)$ by
using their massless expressions.

We keep the $\mHQ$ dependence in the gluon-initiated coefficient
functions ${\mathcal{C}}_{\qHQ/g}$, since those are computed from
the flavor-creation Feynman graphs. The mass-dependent ${\mathcal{O}}(\alpha_{s}/\pi)$
coefficient ${\mathcal{C}}_{\qHQ/g}^{(1)}$ in ${\mathcal{C}}_{\qHQ/g}$
is given by~\cite{Nadolsky:2002jr}\begin{eqnarray}
{\mathcal{C}}_{\qHQ/g}^{(1)}(x,b,\mHQ,\mu_{F}) & = & T_{R}x(1-x)\, b\,\mHQ\, K_{1}(b\,\mHQ)\nonumber \\
 &  & +P_{q/g}^{(1)}(x)\left[K_{0}(b\,\mHQ)-\theta(\mu_{F}-\mHQ)\ln\Bigl(\frac{\mu_{F}}{\mHQ}\Bigr)\right],\label{CHQg}\end{eqnarray}
 where $P_{q/g}^{(1)}(x)=T_{R}\left[x^{2}+(1-x)^{2}\right]$ is the
$q\leftarrow g$ splitting function, $T_{R}=1/2$, and $K_{0}(z)$
and $K_{1}(z)$ are the modified Bessel functions~\cite{AbramowitzStegun}.
The term proportional to the step function $\theta(\mu_{F}-\mHQ)$
is nonzero above the heavy-quark threshold ($\mu_{F}\geq\mHQ$), where
the collinear logarithm $P_{q/g}^{(1)}(x)\ln\left(\mu_{F}/\mHQ\right)$
is subtracted from ${\mathcal{C}}_{\qHQ/g}^{(1)}(x,b,\mHQ,\mu_{F})$
and absorbed into the heavy-quark PDF $f_{\qHQ/A}(x,\mu_{F})$. The
subtraction is not performed below the heavy-quark threshold, because
$f_{\qHQ/A}(x,\mu_{F})=0$ at $\mu_{F}<\mHQ$. For $b\,\mHQ\ll1$,
the gluon-initiated coefficient function reduces to its massless expression,
\begin{equation}
\lim_{b\,\mHQ\rightarrow0}{\mathcal{C}}_{\qHQ/g}^{(1)}(x,b,\mHQ,\mu_{F})=T_{R}x(1-x)-P_{q/g}^{(1)}(x)\ln\left(\mu_{F}b/b_{0}\right).\label{CHQgmeq0}\end{equation}
\begin{figure}[tb]
\begin{center}\includegraphics[%
  clip,
  width=0.47\columnwidth]{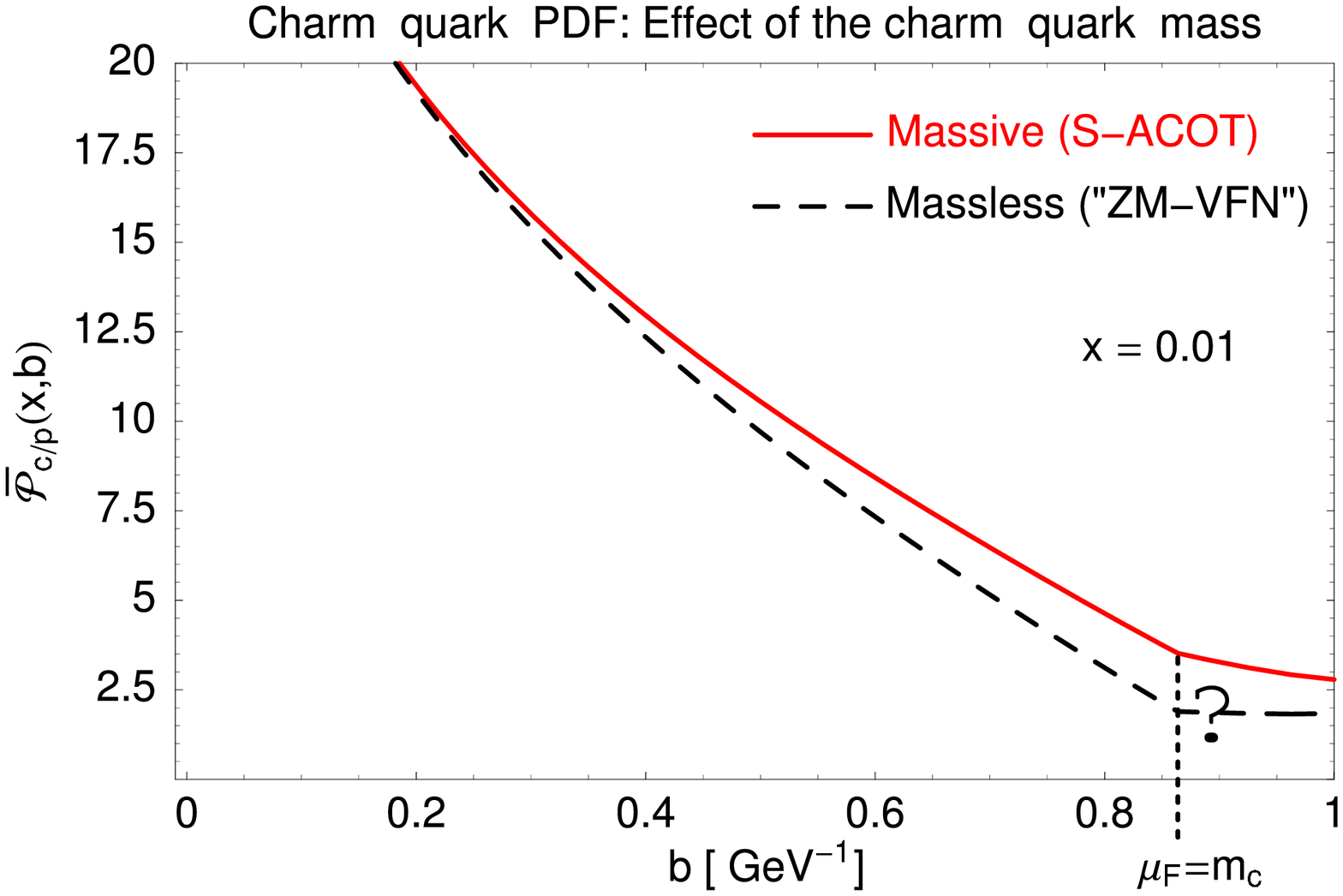}~~~\includegraphics[%
  clip,
  width=0.47\columnwidth]{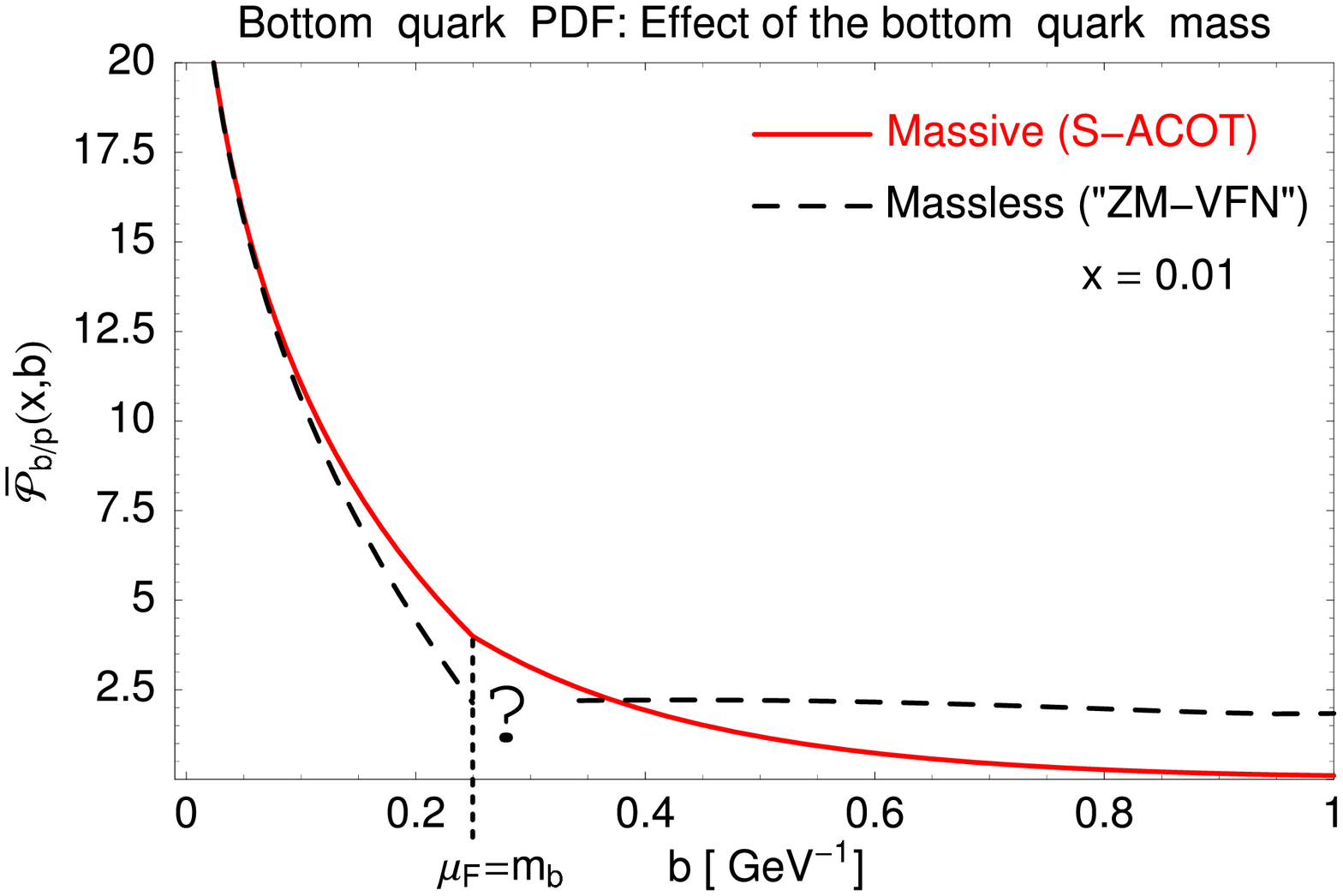}\\
(a)\hspace{7cm}(b)\vspace*{-20pt}\\\end{center}

\caption{The $b$-dependent parton densities $\overline{{\mathcal{P}}}_{\qHQ/A}(x,b,\mHQ)$
vs. the impact parameter $b$ for (a)~charm quarks and (b) bottom
quarks. The solid and dashed curves correspond to the S-ACOT and massless
({}``ZM-VFN'') factorization schemes, respectively.\label{fig:PbA}}
\end{figure}

Convolutions of the Wilson coefficient functions ${\mathcal{C}}_{\qHQ/a}(x,b,\mHQ,\mu_{F})$
and $k_{T}$-integrated PDF's $f_{a/A}(x,\mu_{F})$ are combined into
the $b$-dependent parton density $\overline{{\mathcal{P}}}_{\qHQ/A}(x,b,\mHQ)$
according to Eq.~(\ref{Pja}). At an $n$-th order of $\alpha_{s}$,
$\overline{{\mathcal{P}}}_{\qHQ/A}(x,b,\mHQ)$ and its derivatives
are continuous in $\ln(b)$ up to order $n$, if computed in the S-ACOT
scheme. Here the order of $\overline{{\mathcal{P}}}_{\qHQ/A}(x,b,\mHQ)$
is determined by a formal counting of powers of $\alpha_{s}$ appearing
both in the coefficient functions and PDF's. In this counting, the
PDF $f_{\qHQ/A^{\prime}}(x,\mu_{F})$ for perturbatively generated
heavy quarks is one order higher in $\alpha_{s}$ than the light-parton
PDF's.

At leading order (LO), the function $\overline{{\mathcal{P}}}_{\qHQ/A}(x,b,\mHQ)$
is composed of the heavy-quark-initiated coefficient function ${\mathcal{C}}_{\qHQ/\qHQ}(x,b,\mHQ,\mu_{F})$,
evaluated at order $\alpha_{s}^{0}$, and the gluon coefficient function
${\mathcal{C}}_{\qHQ/g}(x,b,\mHQ,\mu_{F})$, evaluated at order $\alpha_{s}/\pi$:\begin{eqnarray}
\overline{{\mathcal{P}}}_{\qHQ/A}(x,b,\mHQ) & = & \left({\mathcal{C}}_{\qHQ/\qHQ}^{(0)}\otimes f_{\qHQ/A}\right)(x,b,\mHQ,\mu_{F})\nonumber \\
 & + & \frac{\alpha_{s}(\mu_{F})}{\pi}\left({\mathcal{C}}_{\qHQ/g}^{(1)}\otimes f_{g/A}\right)(x,b,\mHQ,\mu_{F})+{\cal O}(\alpha_{s}^{2}/\pi^{2}).\label{PHQALO}\end{eqnarray}
 Here ${\mathcal{C}}_{\qHQ/\qHQ}^{(0)}(x,b,\mHQ,\mu_{F})=\delta(1-x)$.
The dependence of the leading-order (LO) parton density $\overline{{\mathcal{P}}}_{\qHQ/A}(x,b,\mHQ)$
on $b$ is shown for the charm quarks in Fig.~\ref{fig:PbA}(a) and
bottom quarks in Fig.~\ref{fig:PbA}(b). The factorization scale
$\mu_{F}$ is set equal to $b_{0}/b$ to correctly resum the collinear
logarithms in $f_{\qHQ/A}(x,\mu_{F})$ in the small-$b$ limit (cf.~Eq.~(\ref{smallbmHQ})).
Together with the S-ACOT predictions (solid lines), we show the ZM-VFN
predictions (dashed lines).

As discussed above, the ZM-VFN parton density $\overline{{\mathcal{P}}}_{\qHQ/A}(x,b,\mHQ)$
is not properly defined below the threshold $\mu_{F}=\mHQ$ (or above
$b=b_{0}/\mHQ$). We formally define the {}``ZM-VFN'' density $\overline{{\mathcal{P}}}_{\qHQ/A}(x,b,\mHQ)$
at $b>b_{0}/\mHQ$ according to Eq.~(\ref{PHQALO}) by using massless
Wilson coefficient functions ${\mathcal{C}}_{j/a}(x,b,\mHQ=0,\mu_{F})$,
as it was done in the previous resummation calculations. Such a definition
provides just one of many possible continuations of the {}``ZM-VFN''
coefficient functions below the heavy-quark threshold, which render
different results in $q_{T}$ space. We indicate this ambiguity by
enclosing {}``ZM-VFN'' in quotes and placing a question mark by
the dashed line in Fig.~\ref{fig:PbA}. At $b>b_{0}/\mHQ$, the first
term on the right-hand side of Eq.~(\ref{PHQALO}) vanishes ($f_{\qHQ/A}(x,\mu_{F})=0$),
so that $\overline{{\mathcal{P}}}_{\qHQ/A}(x,b,\mHQ)$ becomes equal
to the second term:\begin{eqnarray}
"\mbox{ZM-VFN}",\, b>b_{0}/\mHQ:\nonumber \\
\overline{{\mathcal{P}}}_{\qHQ/A}(x,b,\mHQ=0) & = & \frac{\alpha_{s}(\mu_{F})}{\pi}\left({\mathcal{C}}_{\qHQ/g}^{(1)}\otimes f_{g/A}\right)(x,b,\mHQ=0,\mu_{F})+{\cal O}(\alpha_{s}^{2}/\pi^{2}).\label{ZM-VFNPqHQA}\end{eqnarray}
The massless $\qHQ\leftarrow g$ coefficient function ${\mathcal{C}}_{\qHQ/g}(x,b,\mHQ=0,\mu_{F})$
in Eq.~(\ref{ZM-VFNPqHQA}) is given by Eq.~(\ref{CHQgmeq0}). 

As in the ZM-VFN scheme, the heavy-quark PDF $f_{\qHQ/A}(x,\mu_{F})$
vanishes below the heavy-quark threshold in the S-ACOT scheme. However,
the S-ACOT scheme properly preserves the mass-dependent terms below
the threshold as a part of the gluon-initiated coefficient function
${\mathcal{C}}_{\qHQ/g}^{(1)}(x,b,\mHQ,\mu_{F})$. The S-ACOT parton
density $\overline{{\mathcal{P}}}_{\qHQ/A}(x,b,\mHQ)$ is well-defined
at all $b.$ It reduces to the ZM-VFN result at $b\ll b_{0}/\mHQ$
and is strongly suppressed at $b\gg b_{0}/\mHQ$ by the modified Bessel
functions $K_{0}(b\,\mHQ)$ and $K_{1}(b\,\mHQ)$. The large-$b$
suppression is particularly strong in the case of the bottom quarks,
where $\overline{{\mathcal{P}}}_{b/p}(x,b,m_{b})$ is essentially
negligible at $b>1\mbox{ GeV}^{-1}$ (cf.~Fig.~\ref{fig:PbA}b).
The suppression is caused by the decoupling of the heavy quarks in
the parton densities at $\mu_{F}$ much smaller than $\mHQ$ ($b$
much larger than $b_{0}/\mHQ$). This suppression is independent from
the suppression of contributions with $b\gtrsim0.5-1\mbox{\, GeV}^{-1}$
by the Sudakov exponential $e^{-{\cal S}(b,Q)}$ (see Section~\ref{sec:W-space-form-factor}),
and it does not depend on $Q.$ Consequently the impact of the nonperturbative
contributions from $b\gtrsim1\mbox{\, GeV}^{-1}$ is reduced in the
heavy-quark channels comparatively to the light-quark channels.

\subsection{Nonperturbative contributions\label{sub:Non-perturbative-contributions}}

To gauge the effect of the nonperturbative contributions on the form
factor $\widetilde{W}$, we must estimate the behavior of the Sudakov
exponential $e^{-{\mathcal{S}}(b,Q)}$ and $b$-dependent parton densities
${\mathcal{\overline{P}}}_{j/A}(x,b,\{\mHQ\})$ at large $b$. Many
studies have investigated the nonperturbative contributions to the
resummed form factor; see, e.g., Refs.~\cite{Collins:1982uw,Davies:1984sp,Collins:1985kg,Ladinsky:1994zn,Ellis:1997sc,Landry:1999an,Guffanti:2000ep,Kulesza:2001jc,Kulesza:2002rh,Landry:2002ix,Ji:2004xq,Qiu:2000hf}.
Here we opt to use a new parameterization of the nonperturbative component
found in a global analysis~\cite{KonychevNadolsky:2005xx} of Drell-Yan
pair and $Z$ boson production. This parameterization is obtained
in the revised $b_{*}$~model~\cite{Collins:1982va,Collins:1985kg},
with the free model parameters chosen as to maximally preserve the
exact form of the perturbative contributions at $b<1\mbox{ \, GeV}^{-1}$~\cite{KonychevNadolsky:2005xx}. 

The $b_{*}$~model introduces a function $b_{*}(b,b_{max})\equiv b/\sqrt{1+b^{2}/b_{max}^{2}}$
and defines the resummed form factor $\widetilde{W}$ as\begin{equation}
\widetilde{W}(b,Q,x_{A},x_{B},\{\mHQ\})\equiv\widetilde{W}_{pert}\left(b_{*}(b,b_{max}),Q,x_{A},x_{B},\{\mHQ\}\right)\,\,\, e^{-{\mathcal{F}}_{NP}(b,Q)}\,\label{Wbstar}\end{equation}
at all $b$, with $\widetilde{W}_{pert}\left(b_{*}(b,b_{max}),Q,x_{A},x_{B},\{\mHQ\}\right)$
being the finite-order ({}``perturbative'') approximation to $\widetilde{W}(b,Q,x_{A},x_{B},\{\mHQ\})$.
The higher-order corrections in $\alpha_{s}$ and {}``power-suppressed''
terms proportional to positive powers of $b$ are cumulatively described
by the nonperturbative function ${\mathcal{F}}_{NP}(b,Q)$, defined
as \[
{\mathcal{F}}_{NP}(b,Q)\equiv-\ln\left(\frac{\widetilde{W}(b,Q,x_{A},x_{B},\{\mHQ\})}{\widetilde{W}_{pert}\left(b_{*}(b,b_{max}),Q,x_{A},x_{B},\{\mHQ\}\right)}\right)\]
and found by fitting to the data.

In accordance with Ref.~\cite{KonychevNadolsky:2005xx}, we choose
$b_{max}=1.5\,\mbox{GeV}^{-1}$. We also choose the factorization
scale $\mu_{F}$ in $\left({\mathcal{C}}\otimes f\right)(x,b,\{\mHQ\},\mu_{F})$
such that it stays above the initial scale $Q_{0}\sim1\,$   GeV for
the PDF's $f_{a/A}(x,\mu_{F})$. Specifically, we define $\mu_{F}\equiv b_{0}/b_{*}(b,b_{0}/Q_{0})=(Q_{0}^{2}+b_{0}^{2}/b^{2})^{1/2}$,
so that $\mu_{F}$ is approximately equal to $b_{0}/b$ at $b\rightarrow0$
and asymptotically approaches $Q_{0}$ at $b\rightarrow\infty$.

The function ${\mathcal{F}}_{NP}(b,Q)$ found in the fit of Ref.~\cite{KonychevNadolsky:2005xx}
has the form \begin{equation}
{\mathcal{F}}_{NP}(b,Q)\equiv b^{2}\left[0.201+0.184\,\,\mathrm{ln}\left(Q/(3.2\mbox{\, GeV})\right)-0.026\,\,\mathrm{ln}\left(100\, x_{A}x_{B}\right)\right].\label{KNform}\end{equation}
 This parameterization has several advantages compared to the previous
nonperturbative models. First, it minimizes modifications in $\widetilde{W}_{pert}(b,Q,x_{A},x_{B},\{\mHQ\})$
by the $b_{*}$ prescription in the small-$b$ region, where perturbation
theory is valid. Second, Eq.~(\ref{KNform}) is in a good agreement
with both the available $q_{T}$ data and theoretical expectations.
${\mathcal{F}}_{NP}(b,Q)$ has a quadratic form, ${\mathcal{F}}_{NP}\propto b^{2},$
and follows a nearly linear dependence on $\ln(Q)$, which can be
seen in Eq.~(\ref{KNform}) if the small term $-0.026\ln(100\, x_{A}x_{B})$
is neglected. The quadratic form leads to Gaussian smearing by nonperturbative
{}``intrinsic-$k_{T}$'' effects: $\widetilde{W}(b,Q)\sim e^{-0.5\langle k_{T}^{2}\rangle b^{2}}$,
with $\langle k_{T}^{2}\rangle\lesssim1\mbox{ GeV}^{2}$ in the observable
$Q$ range. Both the quadratic form and linear $\ln(Q)$ dependence
of ${\mathcal{F}}_{NP}(b,Q)$ are suggested by the infrared renormalon
analysis of the large-$b$ contributions~\cite{Korchemsky:1999kt}.
The coefficient $a_{2}=0.184\mbox{\, GeV}^{2}$ of the $\ln(Q)$ term
(found in the fit) agrees well with its independent estimate $\,0.19_{-0.09}^{+0.12}\mbox{\, GeV}^{2}$
obtained within the renormalon analysis~\cite{Tafat:2001in}. The
$\ln(Q)$ term arises from the Sudakov factor as a consequence of
the renormalization-group invariance of the resummed form factor~\cite{Collins:1982va}.
It does not depend on the quark flavors, and it contributes about
70\% to the magnitude of ${\mathcal{F}}_{NP}(b,Q)$ at $Q\sim M_{Z}$.
Therefore, we expect the large flavor-independent term $0.184\, b^{2}\ln(Q/3.2\mbox{\, GeV})$
to be also present in the heavy-quark scattering channels.

The other terms in Eq.~(\ref{KNform}) may in principle depend on
the flavor of the participating quarks. However, no obvious dependence
on the quark flavor or nucleon isospin was observed in the global
$q_{T}$ fit to the Drell-Yan pair and $Z$ boson production data~\cite{KonychevNadolsky:2005xx},
perhaps because these data are mostly sensitive to the scattering
of light $u$ and $d$ quarks. Additional nonperturbative contributions
may be present in the heavy-quark sector, especially if the {}``intrinsic''
heavy-quark states~\cite{Brodsky:1980pb} constitute a non-negligible
part of the proton's wavefunction. In the classical realization, the
intrinsic heavy quarks contribute at large momentum fractions ($x\rightarrow1$),
while the bulk of $W$ and $Z$ boson production happens at much smaller
$x,$ $x=10^{-3}-10^{-1}$. Since the existence of the {}``intrinsic''
heavy quarks remains an open question, we do not consider them in
this study. We therefore parametrize the nonperturbative contributions
by Eq.~(\ref{KNform}), which neglects the flavor dependence. We
vary the parameters of ${\mathcal{F}}_{NP}(b,Q)$ in the heavy-quark
channels in order to evaluate the uncertainty in the physical observables
arising from this approximation.

\subsection{An example of the $b$-space form factor\label{sec:W-space-form-factor}}

\begin{figure}[h]
\begin{center}\includegraphics[%
  width=10cm]{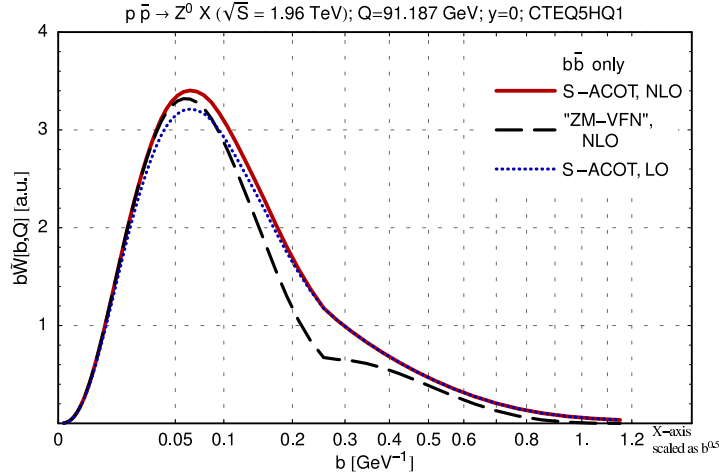}\vspace*{-20pt}\\\end{center}

\caption{The form factor $b\,\widetilde{W}(b,Q)$ vs. the impact parameter
$b$ for $b\bar{b}\rightarrow Z^{0}X$ in the Tevatron Run-2. The
solid, dashed, an dotted curves correspond to the NLO S-ACOT, NLO
{}``ZM-VFN'', and LO S-ACOT calculations, respectively. \label{fig:bWb}}
\end{figure}

Fig.~\ref{fig:bWb}~$\,$shows an example of the full resummed form
factor $b\,\widetilde{W}(b,Q)$ (cf.~Eq.~(\ref{WCSS})) in $Z$
boson production via $b\bar{b}$ annihilation in the Tevatron Run-2.
The form factor is computed in the S-ACOT scheme at the leading-order
(LO) and in the S-ACOT and {}``ZM-VFN'' schemes at the partial next-to-leading-order
(NLO). The leading-order {}``ZM-VFN'' contribution is computed by
using the formal definition of $\overline{{\mathcal{P}}}_{\qHQ/A}(x,b,m_{\qHQ}=0)$
in Eq.~(\ref{PHQALO}). The NLO curves are obtained by adding the
massless flavor-excitation contribution $\left({\mathcal{C}}_{\qHQ/\qHQ}^{(1)}\otimes f_{\qHQ/A}\right)(x,b,\mHQ=0,\mu_{F})$
of order ${\mathcal{O}}(\alpha_{s}/\pi)$ to the LO parton density
$\overline{{\mathcal{P}}}_{\qHQ/A}(x,b,\mHQ)$ in Eq.~(\ref{PHQALO}).
We do not include the ${\mathcal{O}}(\alpha_{s}^{2}/\pi^{2})$ flavor-creation
contribution $\left({\mathcal{C}}_{\qHQ/a}^{(2)}\otimes f_{a/A}\right)(x,b,\mHQ,\mu_{F})$,
with $a=q,\, g$, needed to evaluate $\overline{{\mathcal{P}}}_{\qHQ/A}(x,b,\{\mHQ\})$
at the full NLO accuracy. The partial NLO approximation results in
a discontinuity of the derivative $d\widetilde{W}(b,Q)/d\ln b$ at
the bottom quark threshold, which has little impact on our results.
The perturbative contribution to the Sudakov factor ${\mathcal{S}}(b,Q)$
is evaluated up to order $\alpha_{s}^{2}/\pi^{2}$. The nonperturbative
terms are included according to Eqs.~(\ref{Wbstar}) and~(\ref{KNform}),
assuming flavor independence of ${\mathcal{F}}_{NP}(b,Q)$. Furthermore,
the CTEQ5HQ1 parton distributions~\cite{Lai:1999wy} were used.

According to the figure, the form factor $b\,\widetilde{W}(b,Q)$
in $b\bar{b}\rightarrow Z$ exhibits a maximum at $b\approx0.07\mbox{ GeV}^{-1}$.
Larger impact parameters are suppressed by the Sudakov exponential
$e^{-{\cal S}(b,Q)}$ by an amount growing with $Q$, independently
of the quark flavor and factorization scheme. In the S-ACOT scheme,
the form factor is also suppressed by the $\mHQ$ decoupling in $\overline{{\mathcal{P}}}_{\qHQ/A}(x,b,\mHQ)$.
Consequently the mass-dependent variations in $b\,\widetilde{W}(b,Q)$
are more pronounced in production of not too heavy bosons, such as
$Z^{0}$, and in bottom quark channels, where they occur at relatively
small $b\approx0.25\mbox{ GeV}^{-1}$ (cf.~Fig.~\ref{fig:PbA}). 

The {}``ZM-VFN'' form factor underestimates the S-ACOT form factor
in a wide range of $b$, and it exhibits non-smooth behavior in the
vicinity of the bottom quark threshold. The NLO S-ACOT and {}``ZM-VFN''
form factors agree well at $b<0.05\mbox{\, GeV}^{-1}$. The NLO correction
to $\overline{{\mathcal{P}}}_{\qHQ/A}(x,b,\mHQ)$ mainly contributes
at $b\approx0.02-0.25\mbox{\, GeV}^{-1}$, where it enhances the LO
form factor. The NLO correction is turned off at $b>0.25\mbox{\, GeV}^{-1}$
by the condition $f_{\qHQ/A}(x,\mu_{F})=0$ for $\mu_{F}<\mHQ$. It
is also suppressed at $b<0.02\mbox{\, GeV}^{-1}$ by the small $\alpha_{s}(b_{0}/b)$. 

The differences between the S-ACOT and {}``ZM-VFN'' form factors
seen at $b\approx b_{0}/\mHQ$ will affect the cross sections in $q_{T}$
space at small and moderate transverse momenta.

\section{Numerical results\label{sec:Numerical-results}}

In the present section we compare the resummed $q_{T}$ distributions
calculated in the massive (S-ACOT) and massless ({}``ZM-VFN'') schemes,
defined according to Eq.~(\ref{ZM-VFNPqHQA}). We examine Drell-Yan
production of $W^{\pm}$, $Z^{0}$, and Higgs bosons in the Tevatron
Run-2 (the c.m. energy $\sqrt{S}=1.96$ TeV) and at the LHC ($\sqrt{S}=14$
TeV). In the case of $W$ boson production, we consider the leptonic
decay mode $W\rightarrow e\nu$ and discuss the impact of the different
schemes on the measurement of the $W$ boson mass. We apply the resummation
formalism described in the previous section. The numerical calculation
was performed using the programs Legacy and ResBos~\cite{Balazs:1997xd,Landry:2002ix}
with the CTEQ5HQ1 parton distribution functions~\cite{Lai:1999wy}.
We use the parameterization~(\ref{KNform}) for the nonperturbative
function~${\mathcal{F}}_{NP}(b,Q)$, unless stated otherwise. The
$W^{\pm}$ and $Z^{0}$~boson masses are assumed to be $M_{W}=80.423$~GeV
and $M_{Z}=91.187$~GeV, respectively. The heavy quark masses are
taken to be $m_{c}=1.3$~GeV and $m_{b}=4.5$~GeV.

\subsection{Partonic subprocesses in $W^{\pm}$ and $Z^{0}$ boson production}

We first compare contributions of various partonic subprocesses to
the total $W^{\pm}$ and $Z^{0}$ production cross sections. These
contributions will be classified according to the types of the quarks
$q$ and antiquarks $\bar{q^{\prime}}$ entering the $q\bar{q^{\prime}}W$
or $q\bar{q}Z$ electroweak vertex. At the Born level, the cross section
for production of narrow-width $W^{\pm}$ and $Z^{0}$ bosons in a
$q\bar{q^{\prime}}$ partonic channel is approximately given by\[
\sigma_{q\bar{q^{\prime}}}\approx\sigma_{0}\,\mathcal{L}_{q\bar{q^{\prime}}}(\tau)\, g_{q\bar{q^{\prime}}}\,\,,\]
 where $\tau\equiv Q^{2}/S,$ $Q=M_{W}$ ($M_{Z}$) in $W^{\pm}$
($Z^{0}$) boson production,\begin{equation}
\mathcal{L}_{q\bar{q^{\prime}}}(\tau)=\int_{\tau}^{1}\frac{d\xi}{\xi}\left[f_{q/A}(\xi,Q)\, f_{\bar{q}'/B}(\frac{{\tau}}{\xi},Q)+f_{\bar{q^{\prime}}/A}(\xi,Q)\, f_{q/B}(\frac{{\tau}}{\xi},Q)\right]\end{equation}
is the parton luminosity in the $q\bar{q^{\prime}}$ channel, $g_{q\bar{q^{\prime}}}$
is the flavor-dependent prefactor, and $\sigma_{0}$ includes the
remaining flavor-independent terms \cite{Collins:1985kg}. In $W^{\pm}$
boson production, \begin{equation}
g_{q\bar{q^{\prime}}}=V_{q\bar{q^{\prime}}}^{2},\end{equation}
where $V_{q\bar{q^{\prime}}}$ is the appropriate CKM matrix entry.
In $Z^{0}$ boson production, \begin{equation}
g_{q\bar{q^{\prime}}}=\delta_{qq^{\prime}}\left[(1-4\left|e_{q}\right|\sin^{2}\theta_{w})^{2}+1\right],\end{equation}
where $e_{q}=2/3$ or $-1/3$ is the quark electric charge in units
of the positron charge, and $\theta_{w}$ is the weak mixing angle.
The partial cross sections $\sigma_{q\bar{q^{\prime}}}/\sigma_{tot}$
(where $\sigma_{tot}\equiv\sum_{q,\bar{q^{\prime}}}\sigma_{q\bar{q^{\prime}}}$)
in the different $q\bar{q^{\prime}}$ channels are evaluated as $\mathcal{L}_{q\bar{q^{\prime}}}g_{q\bar{q^{\prime}}}/\left(\sum_{q,\bar{q^{\prime}}}\mathcal{L}_{q\bar{q^{\prime}}}g_{q\bar{q^{\prime}}}\right)$
and listed in Table~\ref{tab:WZfraction}. These values will be modified
somewhat by the NLO radiative corrections included in the following
subsections. 

\begin{table}
\begin{tabular}{|c|>{\centering}m{7mm}|>{\centering}m{7mm}|>{\centering}m{7mm}|>{\centering}m{7mm}|>{\centering}m{7mm}|}
\hline 
&
\multicolumn{5}{c|}{$W^{+}$}\tabularnewline
\hline 
Subprocesses&
$u\bar{d}$&
$u\bar{s}$&
$c\bar{d}$&
$c\bar{s}$&
$c\bar{b}$\tabularnewline
\hline 
Tevatron Run-2&
$90$&
$2$&
$1$&
$7$&
$0$\tabularnewline
\hline 
LHC&
$74$&
$4$&
$1$&
$21$&
$0$\tabularnewline
\hline
\end{tabular}~~\begin{tabular}{|>{\centering}m{7mm}|>{\centering}m{7mm}|>{\centering}m{7mm}|>{\centering}m{7mm}|>{\centering}m{7mm}|}
\hline 
\multicolumn{5}{|c|}{$W^{-}$}\tabularnewline
\hline 
$d\bar{u}$&
$s\bar{u}$&
$d\bar{c}$&
$s\bar{c}$&
$b\bar{c}$\tabularnewline
\hline 
$90$&
$2$&
$1$&
$7$&
$0$\tabularnewline
\hline 
$67$&
$2$&
$3$&
$28$&
$0$\tabularnewline
\hline
\end{tabular}~~\begin{tabular}{|>{\centering}m{7mm}|>{\centering}m{7mm}|>{\centering}m{7mm}|>{\centering}m{7mm}|>{\centering}m{7mm}|}
\hline 
\multicolumn{5}{|c|}{$Z^{0}$}\tabularnewline
\hline 
$u\bar{u}$&
$d\bar{d}$&
$s\bar{s}$&
$c\bar{c}$&
$b\bar{b}$\tabularnewline
\hline 
$57$&
$35$&
$5$&
$2$&
$1$\tabularnewline
\hline 
$36$&
$34$&
$15$&
$9$&
$6$\tabularnewline
\hline
\end{tabular}

\caption{Partial contributions $\sigma_{q\bar{q^{\prime}}}/\sigma_{tot}$
of quark-antiquark annihilation subprocesses to the total Born cross
sections in $W^{\pm}$ and $Z^{0}$ boson production at the Tevatron
and LHC (in percent). \label{tab:WZfraction}\textbf{}}
\end{table}

According to the Table, the contributions of the heavy-quark channels
at the Tevatron are small. In $W$ boson production, only $8\%$ of
the cross section involves heavy (charm) quarks, while light quarks
account for $92\%$. In $Z$ boson production, heavy quarks contribute
even less, with only $3\%$. Due to the small fractional contribution
of the heavy quarks at the Tevatron, the differences between the massive
and the massless calculations are negligible given the expected experimental
uncertainties. 

The LHC probes smaller momentum values $x$, where the fractional
contributions of the initial charm and bottom quarks are much larger.
These channels sum up to $22\%$ in $W^{+}$ boson production, $31\%$
in $W^{-}$ boson production, and $15\%$ in $Z^{0}$ boson production.
Therefore, the following discussion will concentrate on the LHC and
only briefly mention the Tevatron processes.

\subsection{$Z^{0}$ boson production at LHC\label{sub:Numerical-results-Zboson}}

\begin{figure}
\includegraphics[%
  scale=0.4]{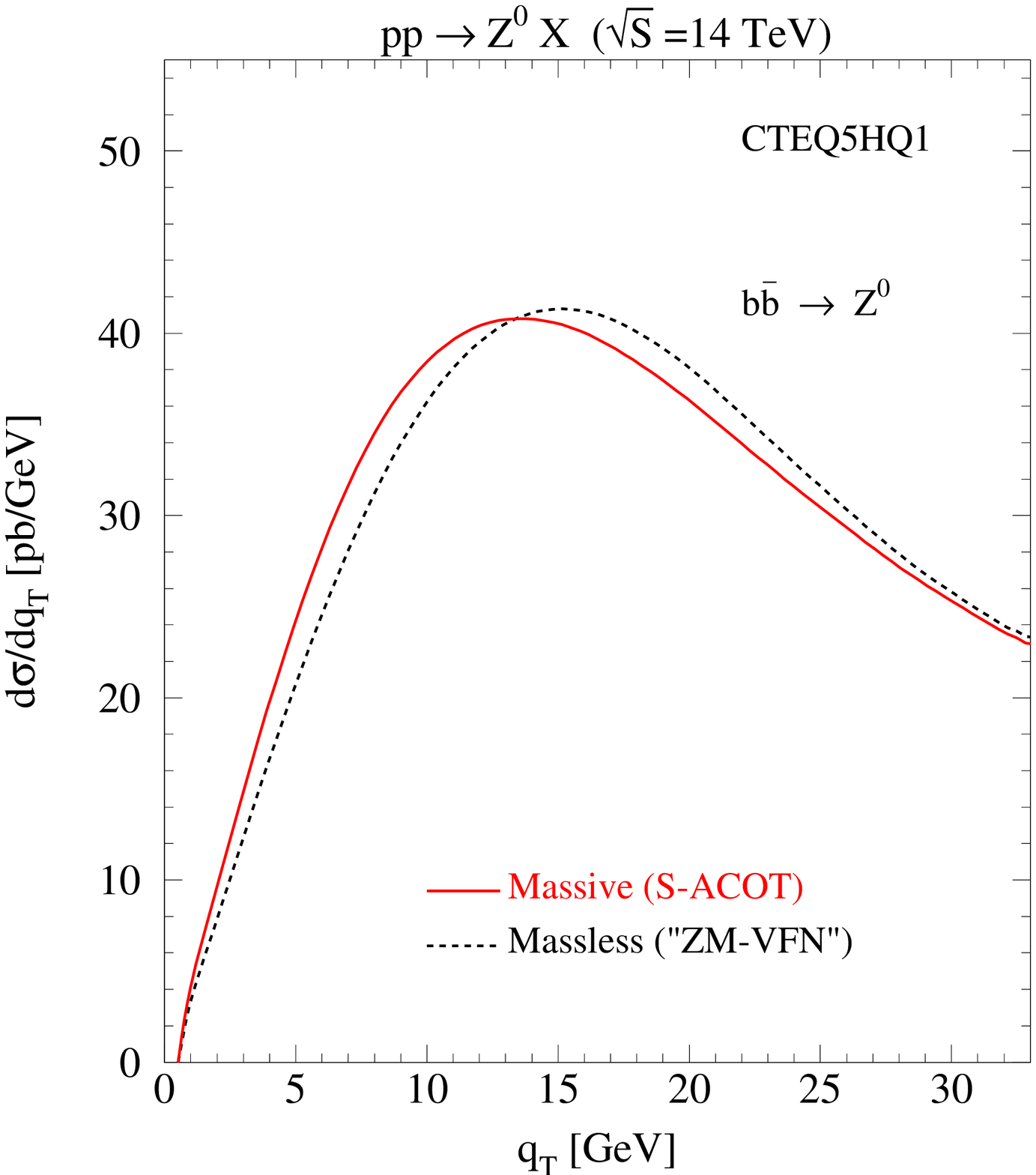}~\hspace*{15pt}\vspace*{-6pt}\includegraphics[%
  scale=0.4]{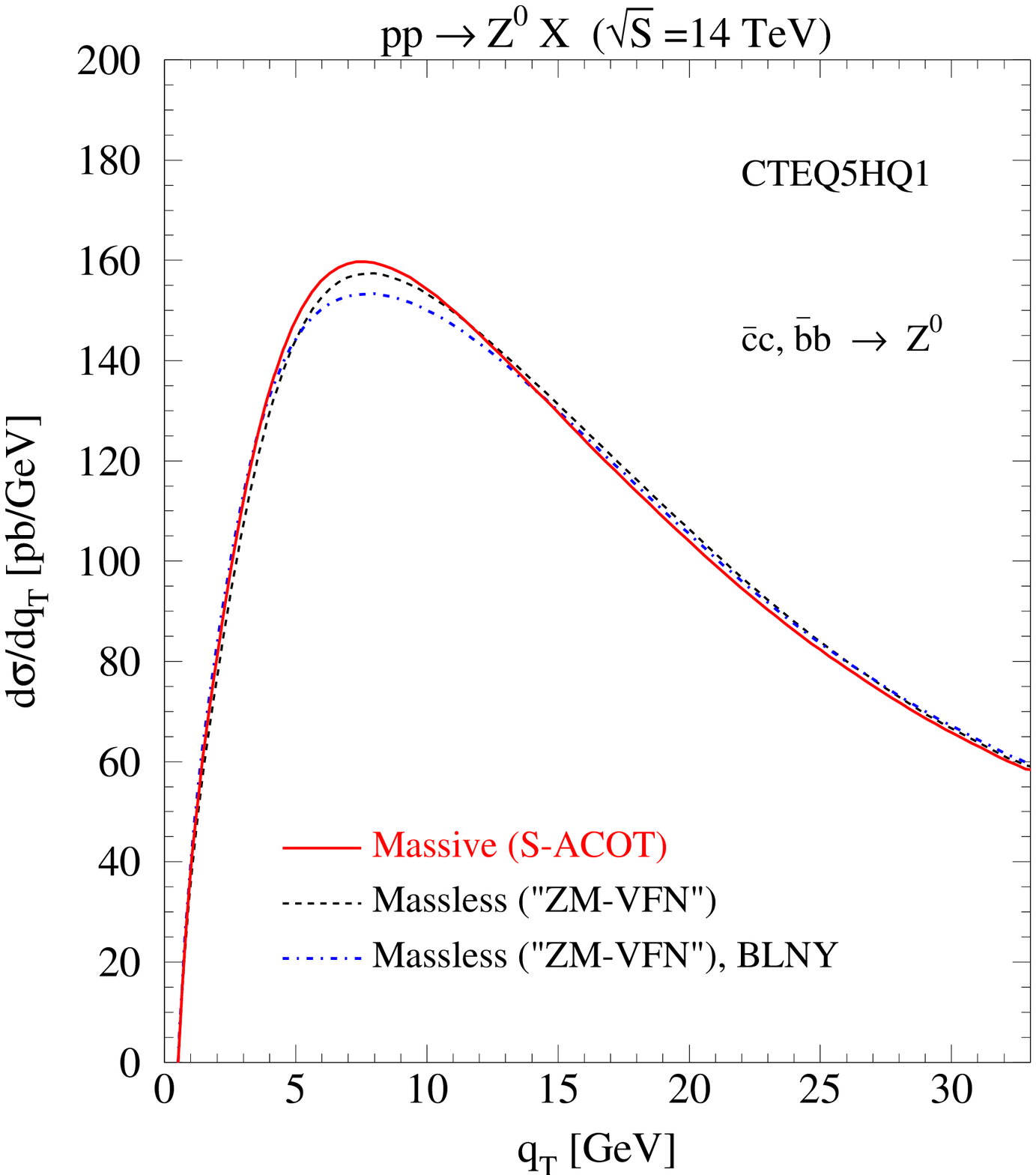}\\
\hspace{0.8cm} (a)\hspace{7.4cm}(b)\vspace*{-10pt}\\

\caption{$d\sigma/dq_{T}$ for $c\bar{c},\, b\bar{b}\to Z^{0}$ boson production
at the LHC: (a) $b\bar{b}$ channel only, (b) combined $c\bar{c}$
and $b\bar{b}$ channels. The solid (red) curve shows the distribution
in the massive (S-ACOT) scheme. The dashed (black) curve shows the
distribution in the massless ({}``ZM-VFN'') scheme, computed using
the parametrization (\ref{KNform}) of the nonperturbative function
${\mathcal{F}}_{NP}(b,Q)$. The dot-dashed (blue) line was calculated
in the {}``ZM-VFN'' scheme using an alternative parameterization~\cite{Landry:2002ix}
of the nonperturbative function~${\mathcal{F}}_{NP}(b,Q)$. \label{fig:ZLHC-Wrap(a)}}
\end{figure}

We first examine the influence of the charm and bottom quark masses
on the resummed $Z^{0}$ boson cross section at the LHC. Fig.~\ref{fig:ZLHC-Wrap(a)}
compares the transverse momentum distribution of $Z^{0}$ boson production
in the heavy-quark ($c\bar{c}$ and $b\bar{b}$) channels at the LHC,
calculated in the S-ACOT and {}``ZM-VFN'' schemes. Fig.~\ref{fig:ZLHC-Wrap(a)}(a)
shows the distribution in the $b\bar{b}$ channel only, where the
effect is the largest due to the large $b$ quark mass. In the approximate
{}``ZM-VFN'' scheme (black dashed line), the $d\sigma/dq_{T}$~distribution
is shifted in the maximum region at $\sim13$~GeV to larger transverse
momenta by about $2$~GeV. This reflects the enhancement of the form
factor $b\,\widetilde{W}(b,Q)$ at $b\sim b_{0}/m_{b}\approx0.25\mbox{ GeV}^{-1}$
in the S-ACOT scheme. The region of large transverse momenta ($q_{T}\gtrsim25$~GeV)
is essentially sensitive only to small impact parameters ($b\lesssim0.1\,\mbox{GeV}^{-1}$),
where the S-ACOT and ZM-VFN form factors are very close. Consequently,
the two schemes give very close predictions in the large-$q_{T}$
region. 

The shift to the larger values of $q_{T}$ is similar in the $c\bar{c}$
channel, albeit smaller due to the lower charm mass. Fig.~\ref{fig:ZLHC-Wrap(a)}(b)
shows the combined distribution in the $c\bar{c}$ and $b\bar{b}$
channels. The mass-induced shift of the peak of the $q_{T}$ distribution
is much smaller, but clearly visible. One should notice that the distribution
of Fig.~\ref{fig:ZLHC-Wrap(a)}(b) peaks around $7.5$~GeV, since
it is dominated by the charm contribution, whereas the bottom distribution
peaks at $15$~GeV (Fig.~\ref{fig:ZLHC-Wrap(a)}(a)). The combined
effect, displayed in Fig.~\ref{fig:ZLHC-Wrap(a)}(b), is an enhancement
of the rate in the peak region of about $2.5\%$.

For comparison, the dot-dashed (blue) line shows the shift of the
{}``ZM-VFN'' $q_{T}$ distribution due to the choice of a different
parameterization for the nonperturbative function~${\mathcal{F}}_{NP}(b,Q)$~\cite{Landry:2002ix}.
The difference between the dotted (black) line and the dot-dashed
(blue) line provides a conservative estimate of the current experimental
uncertainty in ${\mathcal{F}}_{NP}(b,Q)$. The mass-induced effect
is at least comparable to the other uncertainties, when the $c\bar{c}$
and $b\bar{b}$ channels are taken alone. However, these channels
cannot be isolated from the light-quark channels in the single-particle
inclusive observables. According to Table~\ref{tab:WZfraction},
the heavy-quark channels contribute about $15\%$ to the total $Z^{0}$
boson cross section. If the light-quark flavors are included, the
S-ACOT rate is enhanced in the maximum region by a small value of
\textsl{$\sim0.3\%$,} which is beyond the experimental precision
of the LHC.

\subsection{$W^{\pm}$ boson production at LHC\label{sub:Numerical-results-Wboson}}

$W^{\pm}$ boson production at the LHC is of particular interest,
as it will be utilized to measure the $W$~boson mass $M_{W}$ with
high precision (tentatively less than 15 MeV). The $W$~boson mass
can be extracted in the leptonic decay $W\rightarrow e\nu$ by fitting
the theoretical model for different values of $M_{W}$ to kinematical
Jacobian peaks, arising at $M_{T}^{e\nu}\approx M_{W}$ in the leptonic
transverse mass ($M_{T}^{e\nu}$) distribution \cite{Smith:1983aa},
and at $p_{T}^{e}\sim M_{W}/2\approx40$~GeV in the electron $p_{T}^{e}$~spectrum. 

The S-ACOT and {}``ZM-VFN'' schemes yield essentially identical
results for the $M_{T}^{e\nu}$~distribution, suggesting that the
error caused by the massless approximation is likely negligible in
the $M_{T}^{e\nu}$ method. This conclusion follows from the low sensitivity
of the $M_{T}^{e\nu}$ distribution to variations $\delta q_{T}$
in the bosonic $q_{T}$ distributions (suppressed by $\delta q_{T}^{2}/M_{W}^{2}\ll1$).
However, the $p_{T}^{e}$~distribution may be strongly affected.

\begin{figure}
\includegraphics[%
  scale=0.4]{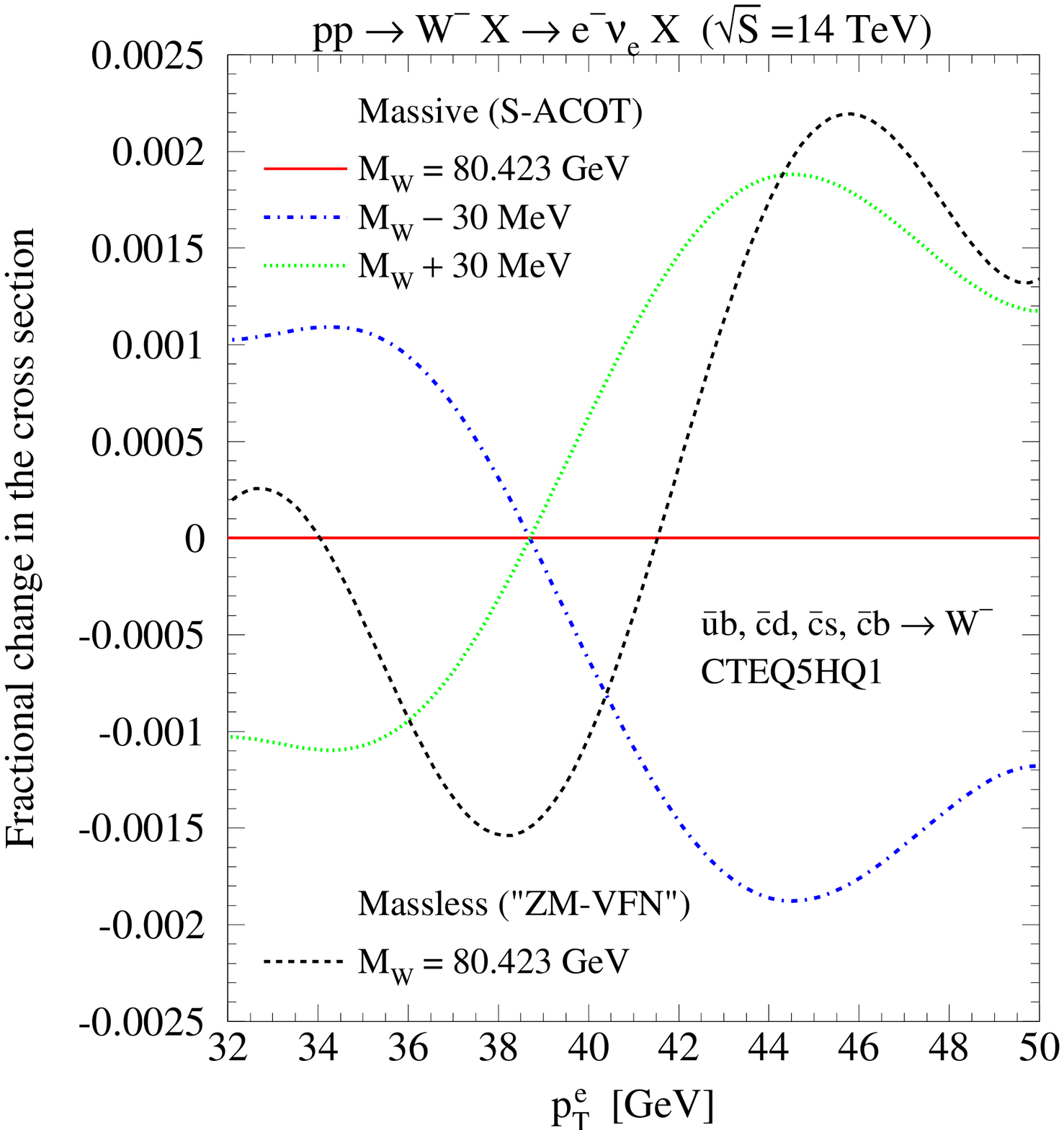}\hspace*{18pt}\includegraphics[%
  scale=0.4]{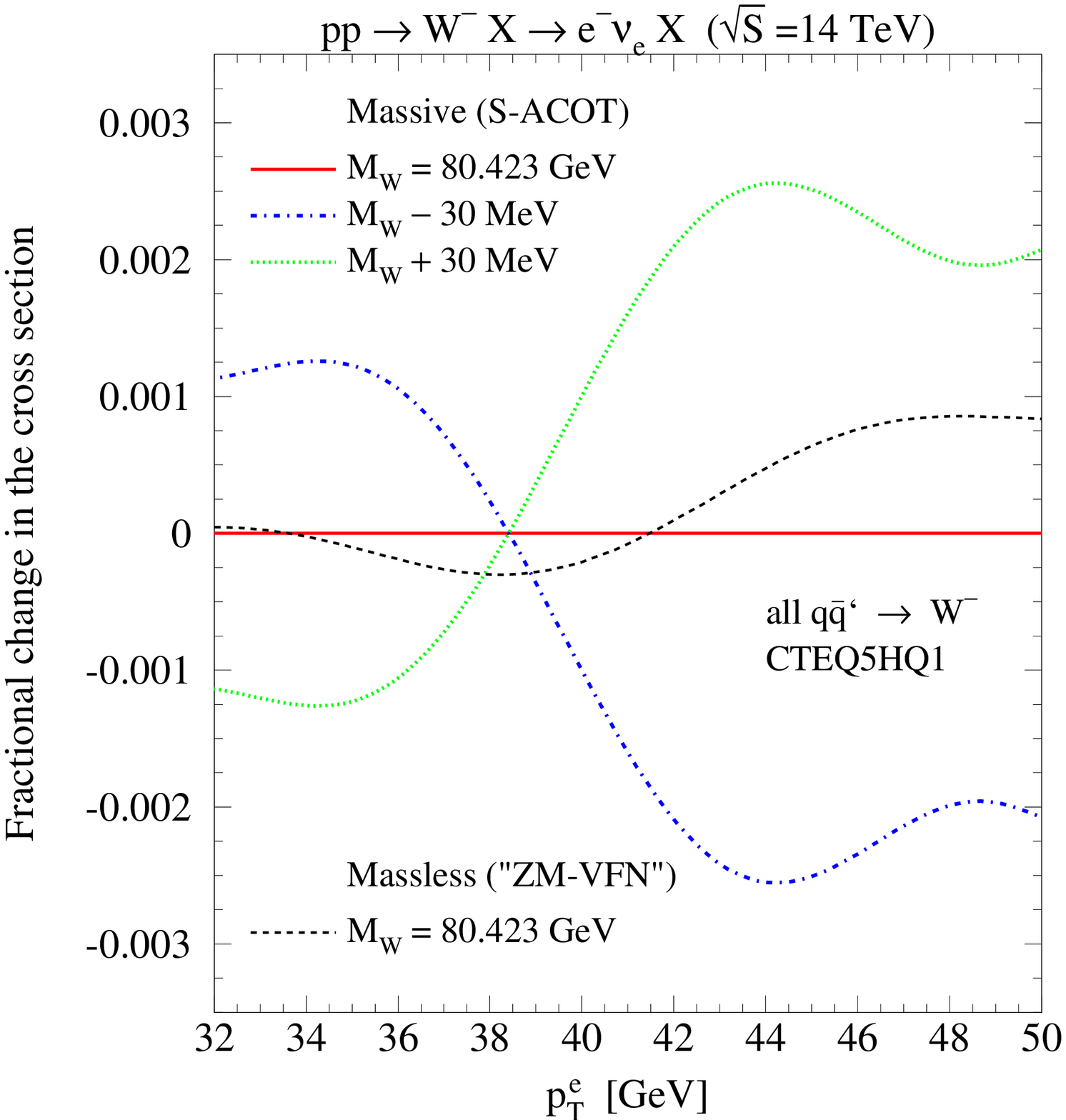}\\
\hspace{1.2cm} (a)\hspace{7.7cm}(b)\vspace*{-10pt}\\

\caption{The fractional change in the distribution $d\sigma/dp_{T}^{e}$ for
$pp\to W^{-}X\to e^{-}\nu_{e}X$ at the LHC: a) only partonic channels
containing heavy quarks ($\bar{u}b,\bar{c}d,\bar{c}s,\bar{c}b\to W^{-}$)
are included; (b)~all channels are included. The dashed (black) line
shows the relative difference between the massless ({}``ZM-VFN'')
and S-ACOT cross sections. The dotted (green) and dot-dashed (blue)
lines show the relative changes in the S-ACOT cross section induced
by variations of $M_{W}$ by $\pm30$~MeV. \label{fig:W-LHC}}
\end{figure}

To better display percent-level changes in $d\sigma/dp_{T}^{e}$ associated
with the effects of the heavy quarks, Fig.~\ref{fig:W-LHC} shows
the fractional difference $\left(d\sigma^{mod}/dp_{T}^{e}\right)/\left(d\sigma^{S-ACOT}/dp_{T}^{e}\right)-1$
of the {}``modified'' (e.g.~{}``ZM-VFN'') cross section and the
{}``standard'' (S-ACOT) cross section. We compare the modifications
due to the approximate massless {}``ZM-VFN'' treatment to the modifications
due to explicit variations of $M_{W}$ in the S-ACOT result.%

We first examine the partonic production channels with at least one
$c$ or $b$ quark (or their antiparticles) in the initial state.
Fig.~\ref{fig:W-LHC}(a) shows the fractional change of the $p_{T}^{e}$
distribution for these channels in $W^{-}$ boson production. The
solid red line shows the reference result, calculated using the S-ACOT
scheme with $M_{W}=80.423$~GeV. The dotted (green) and dot-dashed
(blue) lines show the variation of the S-ACOT $p_{T}^{e}$~distribution,
if the $W$~boson mass is shifted by~$\pm30$~MeV. A small increase
of $M_{W}$ results in a positive shift of the Jacobian peak (dotted
green line), which reduces the cross section at $p_{T}^{e}<M_{W}/2$
and increases the cross section at $p_{T}^{e}>M_{W}/2.$ Smaller values
of $M_{W}$ result in a shift in the opposite direction.

The dashed black curve is the ratio of the $p_{T}^{e}$ distributions
calculated in the {}``ZM-VFN'' and S-ACOT schemes for $M_{W}=80.423$~GeV.
In the heavy-quark channels only (Fig.~(\ref{fig:W-LHC}(a)), the
{}``ZM-VFN'' approximation shifts the Jacobian peak in the positive
direction. The size of this effect is comparable to a shift in $M_{W}$
of about~$+35$~MeV.

Fig.~\ref{fig:W-LHC}(a) is also representative of $W^{+}$~boson
production at the LHC, because the dominant heavy-quark contributions
($c\bar{s}\rightarrow W^{+}$ and $\bar{c}s\rightarrow W^{-}$) are
charge-conjugated, and the parton density functions of $c\,(s)$-quarks
and $\bar{c}\,(\bar{s})$-anti-quarks are almost identical. Hence,
the rates for $W^{+}$ and $W^{-}$ boson production in heavy-quark
scattering are very similar.

\begin{figure}
\begin{center}\hspace*{-2cm}\includegraphics[%
  bb=0bp 0bp 411bp 482bp,
  scale=0.4]{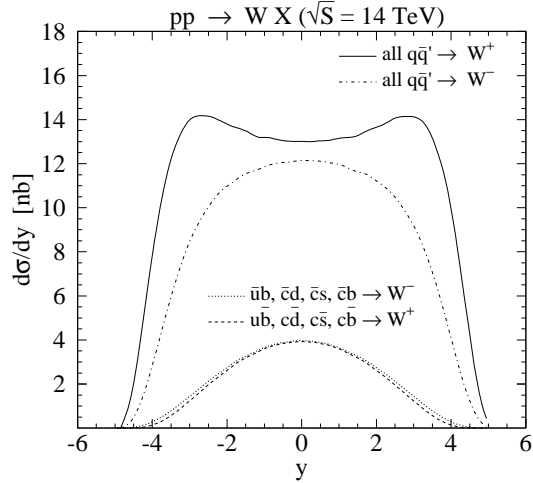}\vspace*{-25pt}\end{center}

\caption{Rapidity distributions of $W^{\pm}$ bosons at the LHC. The solid
(dot-dashed) line is the distribution of the $W^{+}$($W^{-}$) bosons
in all channels. The dashed (dotted) line is the distribution of the
$W^{+}$($W^{-}$) bosons in the channels that involve at least one
initial-state $c$ or $b$ quark.\label{fig:ZLHC-Wrap(b)}}
\end{figure}

Fig.~\ref{fig:W-LHC}(b) shows the fractional change in the $p_{T}^{e}$
distribution for $W^{-}$ bosons, summed over all possible partonic
states.  The total shift of the differential cross section due to
the {}``ZM-VFN'' approximation is comparable with a positive $M_{W}$~shift
of about $10$~MeV. This value is consistent with the shift of $35$
MeV in the $c$ and $b$ channels, in view of the fact that the $c,b$-channels
contribute~$\sim30\%$ to the total $W^{-}$~cross section according
to Table~\ref{tab:WZfraction}. The $q_{T}$ shift is independent
of the rapidity $y$ of the $W^{-}$ boson, because the rapidity distributions
of $W^{-}$ bosons have similar shapes in the heavy-quark and combined
channels (see Fig.~\ref{fig:ZLHC-Wrap(b)}). Consequently the mass-induced
shift in the $W^{-}$ boson $q_{T}$ distribution is not affected
by the rapidity cuts. 

 We contrast this result with $W^{+}$~boson production, where the
profile of the rapidity distribution for the heavy-quark channels
differs from that for the combined channels. The rapidity distribution
of $W^{+}$ bosons in all channels (solid line in Fig.~\ref{fig:ZLHC-Wrap(b)})
has characteristic shoulders at $y\approx\pm3$. These shoulders are
produced by $u\bar{d,}\, u\bar{s}\rightarrow W^{+}$ scattering, enhanced
at large momentum fractions $x$ by the valence $u$-quark distribution.
The fraction of $c$ and $b$ channels (dashed line) is the largest
at central rapidities, $-1<y<1$. In this region, $c$ and $b$ contribute
$\sim29\%$ to the cross section. The difference between the {}``ZM-VFN''
and S-ACOT $p_{T}^{e}$ distributions is comparable here to a change
in $M_{W}$ of about $9$~MeV. In the full rapidity range, this difference
is comparable to $\delta M_{W}\approx6$~MeV. If experimental cuts
appropriate for the ATLAS experiment ($|y_{e}|<2.5$, $p_{T}^{e}>25$~GeV
and $\etmiss>25$~GeV) are applied, the difference is comparable
to $\delta M_{W}\approx7.5$~MeV.

The presented estimates may be modified by additional nonperturbative
contributions arising in the heavy-quark channels, such as those associated
with the {}``intrinsic'' heavy quarks. Our model has neglected such
contributions, in view that their magnitude is uncertain, and they
are less likely to contribute at the small $x$ typical for $W$ and
$Z$ boson production. We assumed that the dominant nonperturbative
contribution at $Q\sim M_{Z}$ ($\approx70\%$ of ${\cal F}_{NP}(b,Q)$)
arises from $Q$ dependence of the Sudakov factor, which does not
depend on the quark flavor (cf.~Section~\ref{sub:Non-perturbative-contributions}).
The {}``ZM-VFN'' and S-ACOT functions differ in $b$ space by several
tens of percent in the threshold region (cf.~ Figs.~\ref{fig:PbA}
and~\ref{fig:bWb}). The extra nonperturbative terms must contribute
at a comparable level to be non-negligible. We confirmed this estimate
numerically by repeating the analysis of the $p_{T}^{e}$ distributions
for a varied (rescaled by a factor of two) function ${\cal F}_{NP}(b,Q)$
in the heavy-quark channels. The resulting variation in $p_{T}^{e}$
distribution for the heavy-quark channels, comparable with $\delta M_{W}$
of a few tens MeV (???), is consistent with the effect of the rescaling
of ${\cal F}_{NP}(b,Q)$ in $b$ space, of order of the difference
of the {}``ZM-VFN'' and S-ACOT form factors $\widetilde{W}(b,Q)$
in the large-$b$ region.

\subsection{$Z^{0}$ and $W^{\pm}$ boson production at the Tevatron}

The dependence on the heavy-quark masses in $W^{\pm}$ and $Z^{0}$
boson production at the Tevatron follows nearly the same pattern as
that at the LHC. In $Z^{0}$ boson production via $c$ and $b$ quark
channels at the Tevatron, the $q_{T}$ distribution in the {}``ZM-VFN''
approximation is shifted toward larger $q_{T}$ with respect to the
S-ACOT distribution, in analogy to the similar shift at the LHC (cf.
Fig.~\ref{fig:ZLHC-Wrap(a)}). The maximum of the $q_{T}$~distribution
in the $b\bar{b}$ channel is shifted by approximately~$2.5$~GeV
if the masses of the quarks are neglected; this is slightly larger
than the shift at the LHC.  However, modifications in the $q_{T}$
distribution in the combined initial-state channels are negligible,
because $c$ and $b$~scattering contributes only~$3\%$ to the
total $Z^{0}$ cross section (cf. Table~\ref{tab:WZfraction}).

The analysis of $W$~boson production at the Tevatron is less involved,
given that the rates for the processes $p\bar{p}\rightarrow W^{+}X$
and $p\bar{p}\rightarrow W^{-}X$ are related by charge conjugation.
The electron $p_{T}^{e}$ distributions in the $c$~and~$b$~quark
channels at the Tevatron qualitatively follow the pattern shown in
Fig.~\ref{fig:W-LHC}(a) for the LHC. The difference between the
$p_{T}^{e}$ distributions in the {}``ZM-VFN'' and S-ACOT schemes
at the Tevatron is comparable to the effect of a $M_{W}$ increase
by~$45$~MeV (vs.~$35$~MeV at the LHC). The heavy-quark channels
contribute just~$8\%$ to the total $W^{\pm}$ cross section at the
Tevatron. Consequently the net $M_{W}$ shift in all channels ($3$~MeV
in the $p_{T}^{e}$ method and $0$ in the $M_{T}^{e\nu}$ method)
is tiny compared to the expected experimental uncertainty of $\sim30-40$~MeV.

\subsection{Higgs boson production\label{sub:bbH}}

In light of our analysis of the $b\bar{b}\to Z^{0}$~channel (Sec.~\ref{sub:Numerical-results-Zboson}),
we anticipate that production of Higgs bosons $H^{0}$ via $b\bar{b}$~annihilation
is also sensitive to the heavy-quark mass effects. In the standard
model, the gluon-gluon fusion process $gg\rightarrow H^{0}$~ is
the dominant production channel for a wide range of Higgs boson masses
at both the Tevatron and LHC~(see, e.g.,~\cite{Assamagan:2004mu,Carena:2002es,Spira:1997dg}
and references therein). Following the gluon-gluon fusion, the next
most important channels at the Tevatron are associated $WH^{0}$ and
$ZH^{0}$ production, and vector boson fusion. At the LHC, the next
largest production modes are the electroweak boson fusion and Higgs-strahlung
processes. The process $b\bar{b}\rightarrow H^{0}$ typically contributes
less than $1\%$ of the total production cross section. In view of
the current uncertainty in the Higgs transverse momentum distribution~\cite{Dobbs:2004bu},
effects due to the heavy quark mass in the $b\bar{b}$ channel will
be negligible in the standard model.

The question becomes more interesting in extensions of the standard
model containing two Higgs doublets, like the minimal supersymmetric
standard model (MSSM), where the Yukawa couplings of the bottom quarks
to the neutral Higgs bosons $h^{0}$, $H^{0}$ and $A^{0}$ depend
on the supersymmetric parameter $\tan\beta$. If $\tan\beta$ is large,
the bottom Yukawa coupling is strongly enhanced, while the top quark
Yukawa coupling remains nearly constant or is suppressed (see Ref.~\cite{Spira:1997dg}).
The $b\bar{b}$~annihilation rate is comparable to the gluon-gluon
fusion rate for medium values of $\tan\beta$ and can dominate the
cross section for $\tan\beta\gtrsim30$.

\begin{figure}
\includegraphics[%
  scale=0.4]{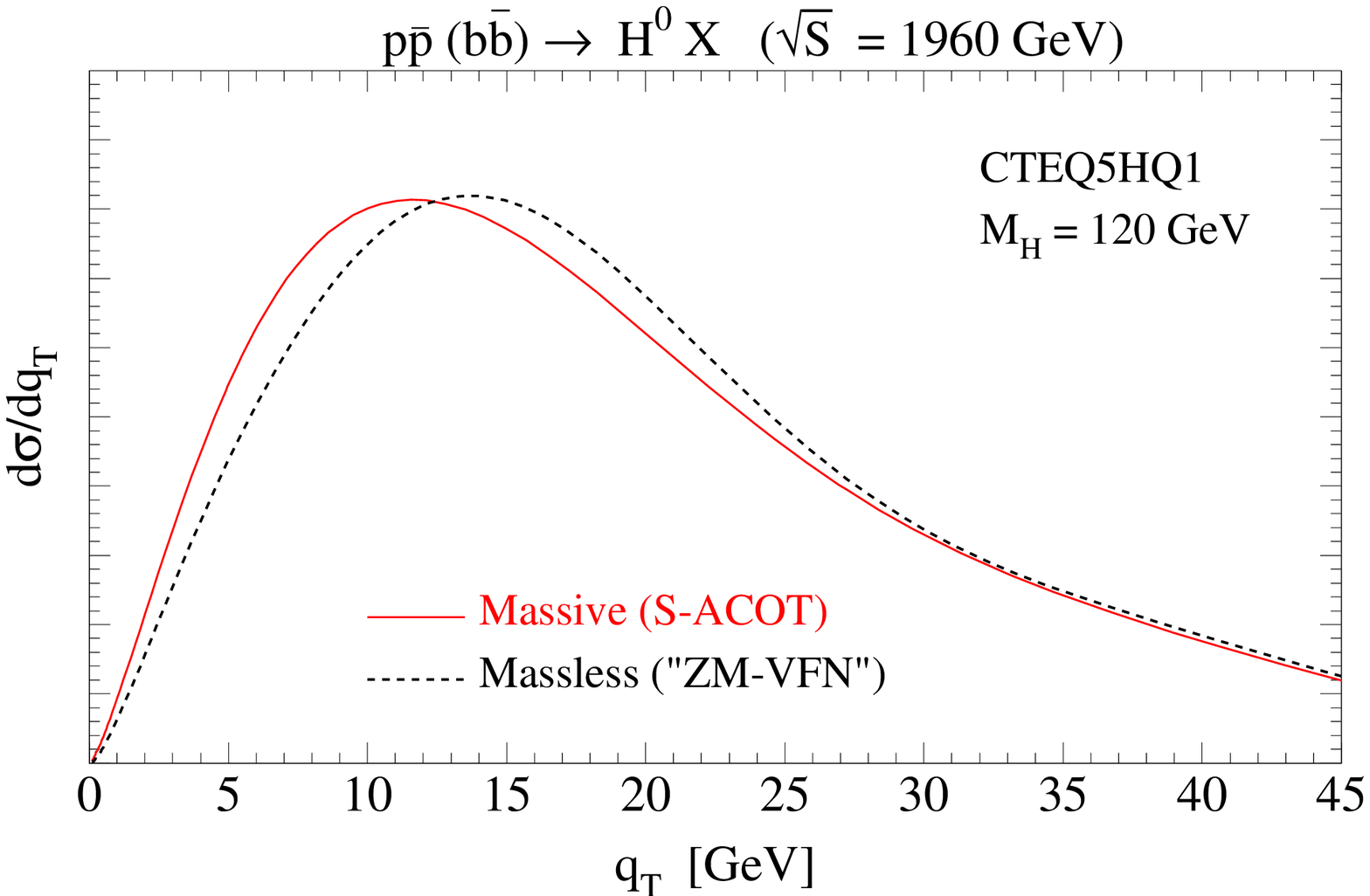}\includegraphics[%
  scale=0.4]{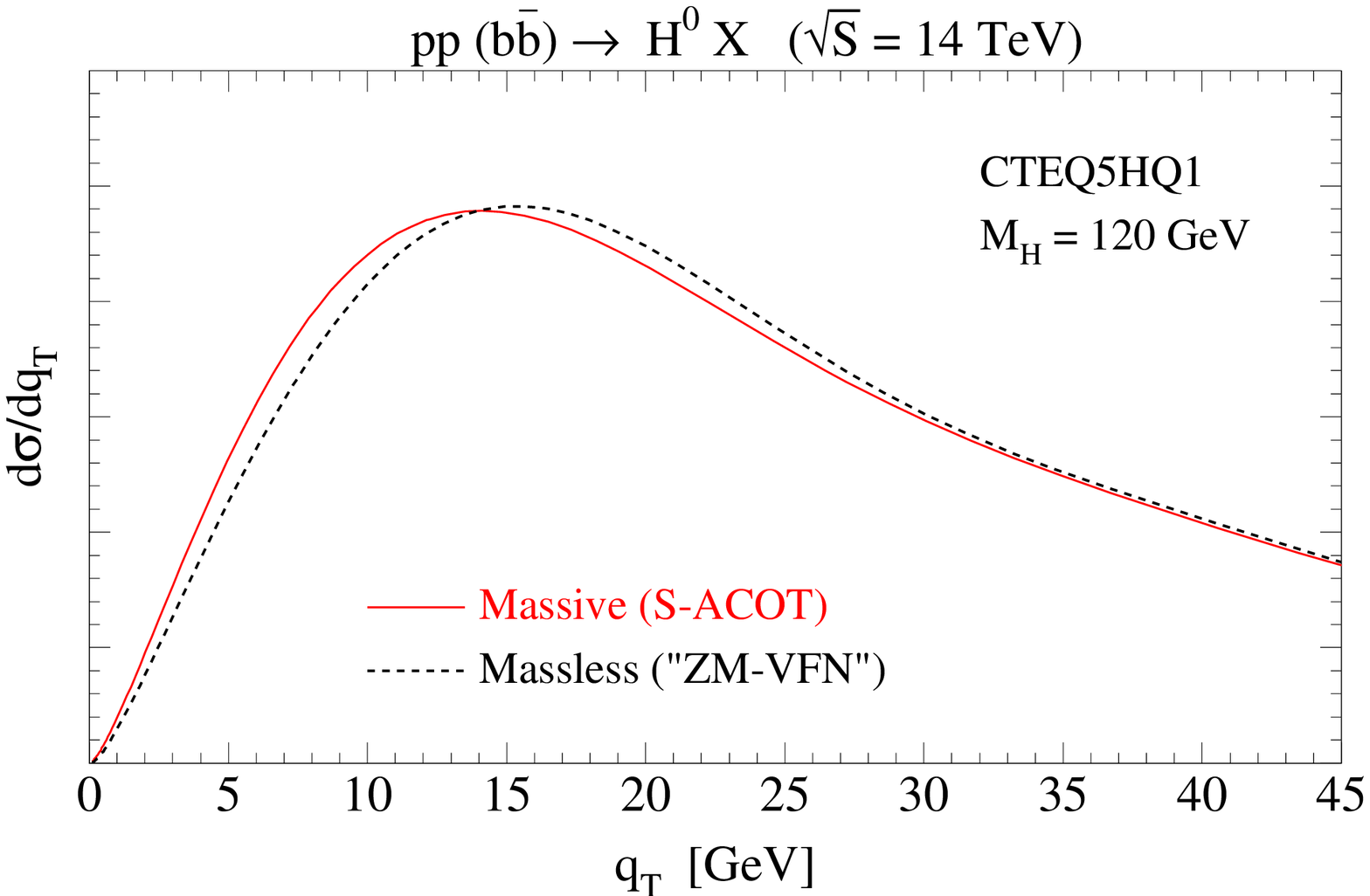}\vspace*{-17pt}

\begin{center}\hspace*{1.cm}(a)\hspace*{7.5cm}(b)\vspace*{-22pt}\end{center}

\caption{Transverse momentum distribution of on-shell Higgs bosons in the
$b\bar{b}\to H^{0}$ channel at (a)~the Tevatron and (b)~LHC. The
solid (red) lines show the $q_{T}$ distribution in the massive (S-ACOT)
scheme. The dashed (black) lines show the distribution in the massless
({}``ZM-VFN'') scheme. \label{fig:bbh_lhc}}
\end{figure}

We implemented the heavy-quark mass effects in the resummation subroutine
for $b\bar{b}\rightarrow H^{0}$ developed in an earlier study~\cite{Balazs:1998sb}.%
\footnote{We thank A. Belyaev and C.-P. Yuan for pointing out a typo in the
${\cal C}_{q/q}$ function in Ref.~\cite{Balazs:1998sb}, which does
not affect the results shown in this section.%
} Fig.~\ref{fig:bbh_lhc} displays $d\sigma/dq_{T}$ for $b\bar{b}\rightarrow H^{0}$~boson
production at (a)~the Tevatron and (b)~LHC. The Higgs boson mass
is chosen to be $M_{H}=120$~GeV. The supersymmetric $b\bar{b}-\mbox{Higgs}$
couplings are obtained at leading order by rescaling the standard
model coupling: $g_{b\bar{b}\{ h^{0},H^{0},A^{0}\}}^{MSSM}=\{-\sin\alpha,\cos\alpha,\sin\beta\,\gamma_{5}\} g_{bbH}^{SM}/\!\cos\beta$~\cite{Gunion:1984yn,Gunion:1989we}.
The net effect of the bottom quark masses on $q_{T}$ distributions
will be the same for both SM and MSSM neutral Higgs bosons, up to
an overall normalization constant. For this reason, Fig.~\ref{fig:bbh_lhc}
does not specify the overall normalization of $q_{T}$ distributions.

The S-ACOT and {}``ZM-VFN'' distributions $d\sigma/dq_{T}$ are
shown by the solid (red) and dashed (black) lines, respectively. A
significant shift of the distribution to larger values of~$q_{T}$
is seen in the {}``ZM-VFN'' approximation. As in $b\bar{b}\to Z^{0}$
process (Sec.~\ref{sub:Numerical-results-Zboson}), the enhancement
of the $q_{T}$~distribution at small $q_{T}$ is caused by the enhancement
in the S-ACOT form factor $b\,\widetilde{W}(b,Q)$ at intermediate
and large $b$ ($b>0.1\,\mbox{GeV}^{-1}$). At the Tevatron, Fig.~\ref{fig:bbh_lhc}(a),
the $q_{T}$ maximum shifts in the {}``ZM-VFN'' approximation to
larger $q_{T}$ by about $2$~GeV out of $11.7$~GeV (about $17\,$\%).
 For a Higgs mass $M_{H}=200$~GeV, the maximum of $d\sigma/dq_{T}$
shifts by about 1.9 GeV out of $12.7$~GeV. 

The corresponding $q_{T}$ distributions for the LHC are shown in
Fig.~\ref{fig:bbh_lhc}(b). The difference between the {}``ZM-VFN''
and S-ACOT calculations is smaller compared to the Tevatron, because
the influence of the region $b>0.1\,\mbox{GeV}^{-1}$ is reduced at
smaller momentum fractions~$x$ probed at the LHC~\cite{Qiu:2000hf}.
The maximum of the transverse momentum distribution shifts in the
{}``ZM-VFN'' approximation by about $1.3$~GeV (9\% out of $14.1$~GeV)
to larger $q_{T}$.  The results for two other Higgs masses ($M_{H}=250$
and $600$ GeV) are summarized in Table~\ref{tab:H_peakshift_LHC}.
The full $q_{T}$ dependence of the $b\bar{b}\rightarrow H$ process
is affected by additional factors, such as the constraints on phase
space available for QCD radiation (less relevant at small $q_{T}$).
The full $q_{T}$ dependence is investigated elsewhere \cite{BelyaevNadolskyYuan}.

\begin{table}[tb]
\begin{center}\begin{tabular}{|>{\centering}p{50mm}|>{\centering}p{20mm}||>{\centering}m{20mm}|>{\centering}m{20mm}|>{\centering}m{18mm}|}
\multicolumn{5}{c}{Differences in $d\sigma/dq_{T}$ for $b\bar{b}\rightarrow H^{0}$
at the LHC}\tabularnewline
\hline 
\multicolumn{2}{|c||}{~$m_{H}$ {[}GeV{]}}&
120&
250&
600\tabularnewline
\hline
\hline 
Position of the&
{}``ZM-VFN''&
15.4&
16.8&
18.8\tabularnewline
maximum {[}GeV{]}&
S-ACOT&
14.1&
15.8&
18.2\tabularnewline
\hline
\hline 
\multicolumn{2}{|c||}{Difference in the positions {[}GeV{]}}&
1.3&
1.0&
0.6\tabularnewline
\hline
\end{tabular}\end{center}

\caption{\label{tab:H_peakshift_LHC} Positions of the maximum of the distribution
$d\sigma/dq_{T}$ in $b\bar{b}\rightarrow H^{0}$ process at the LHC
in the mass-dependent (S-ACOT) and massless ({}``ZM-VFN'') calculations,
as well as the differences of these positions. }
\end{table}

\section{Conclusion\label{sec:Conclusion}}

In the present paper, we make use of recent theoretical developments
to estimate the impact of heavy-quark masses $\mHQ$ on transverse
momentum ($q_{T}$) distributions in production of $W,$ $Z,$ and
Higgs bosons at the Tevatron and LHC. We note that the the zero-mass
({}``massless'') variable flavor number scheme is not consistent
in the heavy-quark channels, if $q_{T}$ is of order of the heavy-quark
mass $\mHQ$. To properly assess the $\mHQ$ dependence, we perform
resummation of the large logarithms $\ln(q_{T}^{2}/Q^{2})$ in the
Collins-Soper-Sterman approach, formulated in a general variable flavor
number factorization scheme (S-ACOT scheme) to correctly resum the
collinear logarithms $\ln(\mu_{F}/\mHQ)$. 

We compare our consistent treatment of the heavy-quark mass dependence
with an approach based on the ZM-VFN scheme. The proper treatment
of the heavy-quark masses leads to an enhancement of the form factor
$\widetilde{W}(b,Q)$ at intermediate and large impact parameters,
which in turn increases the transverse momentum distribution at small
~$q_{T}$. In Drell-Yan production of heavy bosons, the cumulative
effect in the S-ACOT scheme shifts the peak of the $d\sigma/dq_{T}$
distribution to smaller values of~$q_{T}$. The mass effects are
negligible if $q_{T}$ is of order $Q$.

The mass dependence is magnified in subprocesses dominated by scattering
of initial-state bottom quarks, as a consequence of the larger $b$-quark
mass. In the production of $Z^{0}$ and light Higgs bosons via $b\bar{b}$
annihilation at the LHC, the maximum of $d\sigma/dq_{T}$ shifts from
$13.5-14$ GeV in the S-ACOT scheme to $\approx15$ GeV in the {}``ZM-VFN''
approximation. At the Tevatron, the maximum shifts from $\approx10$
to $12$ ~GeV. Our computation is applied to single-particle inclusive
production, when no heavy quarks are observed in the final state.
Since the single-particle inclusive cross sections receive contributions
from both the heavy-quark channels and intensive light-quark channels,
the $\mHQ$ dependence is reduced below the experimental sensitivity
in several cases, notably in $W^{\pm}$ and $Z^{0}$ boson production
at the Tevatron, and $Z^{0}$ boson production at the LHC. On the
other hand, the signal from the heavy-quark channels may be enhanced
by tagging one or two heavy quarks in the final state. In that case,
mass-dependent effects of a similar magnitude may be observed in some
regions of phase space, \emph{e.g.,} at small $q_{T}$ of the boson-heavy
quark system.

Since the LHC plans to measure the $W$ boson mass $M_{W}$ to a high
precision, we must fully account for all potential uncertainties,
including the heavy-quark mass effects. At the LHC, the massless {}``ZM-VFN''
approximation introduces a bias in the measurement of $M_{W}$ from
the transverse momentum distribution $d\sigma/dp_{T}^{e}$ of the
electrons, comparable to a positive $M_{W}$ shift by $\sim10$~MeV.
In $W^{+}$ boson production the bias is comparable to the $M_{W}$
shift of $9$~MeV at central rapidities of $W$ bosons ($|\, y\,|\leq1$)
and $6$~MeV in the whole rapidity range. These effects are not negligible
in view of the desired precision of the $M_{W}$ measurement of $\sim15$~MeV.

\section*{Acknowledgments}

We thank A.~Belyaev and C.-P.~Yuan for valuable discussions. An
independent study \cite{BelyaevNadolskyYuan} of the resummation in
the $b\bar{b}\rightarrow\mbox{Higgs}$ process, including the heavy-quark
mass effects, has been brought to our attention as we completed this
paper. We also thank U.~Baur and S.~Heinemeyer for helpful communications.
F.$\,$I.$\,$O. and S.$\,$B. acknowledge the hospitality of Fermilab
and BNL, where a portion of this work was performed. The work of S.
B. and F. I. O. was partially supported by the U.S. Department of
Energy under grant DE-FG03-95ER40908 and the Lightner-Sams Foundation.
The work of P. M. N. was supported in part by the U.S. Department
of Energy, High Energy Physics Division, under Contract W-31-109-ENG-38. 

\bibliographystyle{apsrev}
\bibliography{bibolness4}

\end{document}